\newif\ifapj
\apjtrue

\ifapj
  \documentclass[twocolumn]{aastex6}
\else
  \documentclass[12pt,preprint]{aastex}
\fi

\usepackage{amsmath,apjfonts,amsfonts,amssymb,enumerate,graphicx,
  longtable,float,times,textcomp,verbatim,url,nicefrac}
\newcommand{\B}{{\it B}-band}
\newcommand{\V}{{\it V}-band}
\newcommand{\R}{{\it R}-band}
\newcommand{\I}{{\it I}-band}

\newcommand{\usdss}{{\it u}-band}
\newcommand{\g}{{\it g}-band}
\newcommand{\r}{{\it r}-band}
\newcommand{\isdss}{{\it i}-band}
\newcommand{\z}{{\it z}-band}

\newcommand{\angstrom}{\text{\normalfont\AA}}

\tabletypesize{\footnotesize}
\shorttitle{Accretion Disk Lags in 2 AGN} 
 \shortauthors{Fausnaugh et  al.}

\begin{document}

\title{Continuum Reverberation Mapping of the Accretion Disks in Two Seyfert 1
  Galaxies}

\author{
  M.~M.~Fausnaugh\altaffilmark{1,2},
  D.~A.~Starkey\altaffilmark{3},
  Keith~Horne\altaffilmark{3},
  C.~S.~Kochanek\altaffilmark{1,4},
  B.~M.~Peterson\altaffilmark{1,4,5},
  M.~C.~Bentz\altaffilmark{6},
  K.~D.~Denney\altaffilmark{7},
  C.~J.~Grier\altaffilmark{8,9},
  D.~Grupe\altaffilmark{10},
  R.~W.~Pogge\altaffilmark{1,4},
  G.~De Rosa\altaffilmark{5},
  S.~M.~Adams\altaffilmark{11},
  A.~J.~Barth\altaffilmark{12},
  Thomas~G.~Beatty\altaffilmark{8,13},
  A.~Bhattacharjee\altaffilmark{14,15}
  G.~A.~Borman\altaffilmark{16},
  T.~A.~Boroson\altaffilmark{17},
  M.~C.~Bottorff\altaffilmark{18},
  Jacob~E.~Brown\altaffilmark{19},
  Jonathan~S.~Brown\altaffilmark{1},
  M.~S.~Brotherton\altaffilmark{14},
  C.~T.~Coker\altaffilmark{1},
  S.~M.~Crawford\altaffilmark{20},
  K.V.~Croxall\altaffilmark{7},
  Sarah~Eftekharzadeh\altaffilmark{14},
  Michael~Eracleous\altaffilmark{8,9,21},
  M.~D.~Joner\altaffilmark{22},
  C.~B.~Henderson\altaffilmark{23,24},
  T.~W.-S.~Holoien\altaffilmark{1,4},
  T.~Hutchison\altaffilmark{18},
  Shai~Kaspi\altaffilmark{25},
  S.~Kim\altaffilmark{1},
  Anthea~L.~King\altaffilmark{26},
  Miao~Li\altaffilmark{27},
  Cassandra~Lochhaas\altaffilmark{1},
  Zhiyuan~Ma\altaffilmark{19},
  F.~MacInnis\altaffilmark{18},
  E.~R.~Manne-Nicholas\altaffilmark{5},
  M.~Mason\altaffilmark{14},
  Carmen~Montuori\altaffilmark{28},
  Ana~Mosquera\altaffilmark{29},
  Dale~Mudd\altaffilmark{1},
  R.~Musso\altaffilmark{18},
  S.~V.~Nazarov\altaffilmark{16},
  M.~L.~Nguyen\altaffilmark{14},
  D.~N.~Okhmat\altaffilmark{16},
  Christopher A. Onken\altaffilmark{30},
  B.~Ou-Yang\altaffilmark{6},
  A.~Pancoast\altaffilmark{31,32},
  L.~Pei\altaffilmark{12,33},
  Matthew~T.~Penny\altaffilmark{1,34},
  Rados\l{}aw~Poleski\altaffilmark{1},
  Stephen~Rafter\altaffilmark{35},
  E.~Romero-Colmenero\altaffilmark{19,36},
  Jessie~Runnoe\altaffilmark{8,9,37},
  David~J.~Sand\altaffilmark{38},
  Jaderson~S.~Schimoia\altaffilmark{39},
  S.~G.~Sergeev\altaffilmark{16},
  B.J.~Shappee\altaffilmark{40,41,42},
  Gregory~V.~Simonian\altaffilmark{1},
  Garrett~Somers\altaffilmark{43,44},
  M.~Spencer\altaffilmark{22},
  Daniel~J.~Stevens\altaffilmark{1},
  Jamie~Tayar\altaffilmark{1},
  T.~Treu\altaffilmark{45},
  Stefano~Valenti\altaffilmark{46}, 
  J.~Van Saders\altaffilmark{40,41},
  S.~Villanueva Jr.\altaffilmark{1},
  C.~Villforth\altaffilmark{47},
  Yaniv~Weiss\altaffilmark{48},
  H.~Winkler\altaffilmark{49},
  W.~Zhu\altaffilmark{1}
}

\altaffiltext{1}{Department of Astronomy, The Ohio State University,
  140 W 18th Ave, Columbus, OH 43210, USA}

\altaffiltext{2}{MIT Kavli Institute for Astrophysics and Space Research,
  77 Massachusetts Avenue, 37-241, Cambridge, MA 02139, USA}

\altaffiltext{3}{SUPA Physics and Astronomy, University of
St. Andrews, Fife, KY16 9SS Scotland, UK}

\altaffiltext{4}{Center for Cosmology and AstroParticle Physics, The
  Ohio State University, 191 West Woodruff Ave, Columbus, OH 43210,
  USA}

\altaffiltext{5}{Space Telescope Science Institute, 3700 San Martin Drive, Baltimore, MD 21218, USA}

\altaffiltext{6}{Department of Physics and Astronomy, Georgia State University, Atlanta,
GA 30303}

\altaffiltext{7}{Illumination Works LLC., 5650 Blazer Parkway, Suite
  152 Dublin, OH 43017}


\altaffiltext{8}{Department
  of Astronomy \& Astrophysics, The Pennsylvania State University, 525
  Davey Laboratory, University Park, PA 16802, USA}
\altaffiltext{9}{Institute for Gravitation and the Cosmos, The Pennsylvania State University, University Park, PA 16802, USA}

\altaffiltext{10}{Department of Earth and Space Sciences, Morehead State University, Morehead, Kentucky, USA}

\altaffiltext{11}{Cahill Center for Astrophysics, California Institute of Technology, Pasadena, CA 91125, USA}

\altaffiltext{12}{Department of Physics and Astronomy, 4129 Frederick
Reines Hall, University of California, Irvine, CA 92697, USA}

\altaffiltext{13}{Center for Exoplanets and Habitable Worlds, The Pennsylvania State University, 525 Davey Lab, University Park, PA 16802}

\altaffiltext{14}{Department of Physics and Astronomy, University of
Wyoming, 1000 E. University Ave. Laramie, WY, USA}

\altaffiltext{15}{The Department of Biology, Geology, and Physical Sciences, Sul Ross State
University, WSB 216, Box-64, Alpine, TX, 79832, USA}

\altaffiltext{16}{Crimean Astrophysical Observatory, P/O Nauchny,
Crimea 298409, Russia}

\altaffiltext{17}{Las Cumbres Global Telescope Network, 6740 Cortona Drive, Suite 102,
Santa Barbara, CA  93117, USA}

\altaffiltext{18}{Fountainwood Observatory, Department of Physics FJS 149,
Southwestern University, 1011 E. University Ave., Georgetown, TX 78626, USA}

\altaffiltext{19}{Department of Physics and Astronomy, University of Missouri, Columbia, USA}

\altaffiltext{20}{South African Astronomical Observatory, P.O. Box
  9, Observatory 7935, Cape Town, South Africa}

\altaffiltext{21}{Center for Relativistic Astrophysics, Georgia Institute of Technology, Atlanta, GA 30332, USA}

\altaffiltext{22}{Department of Physics and Astronomy, N283 ESC, Brigham Young University,
Provo, UT 84602-4360, USA}

\altaffiltext{23}{Jet Propulsion Laboratory, California Institute of Technology, 4800 Oak Grove Drive, Pasadena, CA 91109, USA}
\footnotetext[24]{NASA Postdoctoral Program Fellow}

\altaffiltext{25}{School of Physics and Astronomy, Raymond and Beverly Sackler Faculty of Exact
Sciences, Tel Aviv University, Tel Aviv 69978, Israel}

\altaffiltext{26}{School of Physics, University of Melbourne, Parkville, VIC 3010, Australia}

\altaffiltext{27}{Department of Astronomy, Columbia University, 550 W120th Street, New York, NY 10027, USA}
\altaffiltext{28}{DiSAT, Universita dell'Insubria, via Valleggio 11, 22100, Como, Italy}

\altaffiltext{29}{Physics Department, United States Naval Academy, Annapolis, MD 21403}

\altaffiltext{30}{ Research School of Astronomy and Astrophysics, Australian National University, Canberra, ACT 2611, Australia}

\altaffiltext{31}{Harvard-Smithsonian Center for Astrophysics, 60 Garden Street, Cambridge, MA 02138, USA}
\footnotetext[32]{Einstein Fellow}

\altaffiltext{33}{Department of Astronomy, University of Illinois at Urbana-
Champaign, Urbana, IL 61801, USA}

\footnotetext[34]{Sagan Fellow}
\altaffiltext{35}{Department of Physics, Faculty of Natural Sciences, University of Haifa,
   Haifa 31905, Israel}

\altaffiltext{36}{Southern African Large Telescope Foundation,
  P.O. Box 9, Observatory 7935, Cape Town, South Africa}

\altaffiltext{37}{Department of Astronomy, University of Michigan,
  1085 S. University, 311 West Hall, Ann Arbor, MI 48109-1107}

\altaffiltext{38}{Physics \& Astronomy Department, Texas Tech University, Box 41051, Lubbock, TX 79409-1051, USA}

\altaffiltext{39}{Instituto de F\'isica, Universidade Federal do Rio Grande do Sul, Campus do Vale, Porto Alegre, RS, Brazil}

\altaffiltext{40}{Carnegie Observatories, 813 Santa Barbara Street, Pasadena, CA 91101, USA}
\footnotetext[41]{Carnegie-Princeton Fellow}
\footnotetext[42]{Hubble Fellow}
\altaffiltext{43}{Department of Physics and Astronomy, Vanderbilt University, 6301 Stevenson Circle, Nashville, TN, 37235}
\altaffiltext{44}{VIDA Postdoctoral Fellow}

\altaffiltext{45}{Department of Physics and Astronomy, University of
California, Los Angeles, CA 90095-1547, USA}

\altaffiltext{46}{Department of Physics, University of California, One
  Shields Avenue, Davis, CA 95616, USA}

\altaffiltext{47}{University of Bath, Department of Physics, Claverton Down, BA2 7AY, Bath, United Kingdom}

\altaffiltext{48}{Physics Department, Technion, Haifa 32000, Israel}

\altaffiltext{49}{Department of Physics, University of Johannesburg,
  PO Box 524, 2006 Auckland Park, South Africa}

\begin{abstract}
  We present optical continuum lags for two Seyfert 1 galaxies,
  MCG+08-11-011 and NGC\,2617, using monitoring data from a
  reverberation mapping campaign carried out in 2014.  Our light
  curves span the {\it ugriz} filters over four months, with median
  cadences of 1.0 and 0.6 days for MCG+08-11-011 and NGC\,2617,
  respectively, combined with roughly daily X-ray and near-UV data
  from {\it Swift} for NGC 2617.  We find lags consistent with
  geometrically thin accretion-disk models that predict a
  lag-wavelength relation of $\tau \propto \lambda^{4/3}$.  However,
  the observed lags are larger than predictions based on standard
  thin-disk theory by factors of 3.3 for MCG+08-11-011 and 2.3 for
  NGC\,2617.  These differences can be explained if the mass accretion
  rates are larger than inferred from the optical luminosity by a
  factor of 4.3 in MCG+08-11-011 and a factor of 1.3 in NGC\,2617,
  although uncertainty in the SMBH masses determines the significance
  of this result.  While the X-ray variability in NGC\,2617 precedes
  the UV/optical variability, the long 2.6 day lag is problematic for
  coronal reprocessing models.
\end{abstract}

\section{Introduction}
Energy generation in active galactic nuclei (AGN) is believed to be
due to an accretion disk around a super-massive black hole (SMBH).
Viscous torques in the disk caused by magnetic fields move matter
closer to the SMBH and convert gravitational potential energy into
heat and radiation (e.g., \citealt{Page1974, Rees1984, Balbus1998}).
The disk reaches $10^5$--$10^6$\,K at its inner edge with a gradient
to cooler temperatures at larger radii, leading to a continuum
emission spectrum spanning the extreme ultraviolet (UV) to the
infrared (IR).  This model is sufficient to explain the large
luminosities and UV peaks of typical AGN spectral energy distributions
\citep{Burbidge1967, Weedman1977, Sheilds1978, Elvis1994, Telfer2002}.
The UV/optical continuum from the disk also provides the seed photons
that are reprocessed into the IR by hot dust (e.g.,
\citealt{Suganuma2006,Nenkova2008, Vazquez2015}) and X-rays by a
putative hot ``corona'' (e.g., \citealt{Haardt1991, Reynolds2003,
  Turner2006}).  The hottest parts of the disk supply the ionizing
photons that power Doppler-broadened emission lines in the broad- and
narrow-line regions (BLRs and NLRs, \citealt{Davidson1979,
  Veilleux1987}).

Many important aspects of the accretion disk and the continuum
emission remain unknown.  On the theoretical side, there are several
viable accretion-disk models, ranging from geometrically thin disks
that radiate thermally \citep{Shakura1973}, to thick toroids that
radiate through plasma processes \citep{Abramowicz1988} and
radiatively inefficient accretion flows that advect most of their
energy across the event horizon \citep{Narayan1995}.  It is also
challenging to account for the X-ray-emitting corona from first
principles \citep{Schnittman2013a}, and there are a wide variety of
hypothesized geometries and energetic connections between the corona
and the accretion disk.  Simulations have recently made progress by
incorporating radiation transport (e.g., \citealt{Schnittman2013b,
  Sadowski2014, Sadowski2015a}), but it has not been possible to
simulate the full magneto-radiation-hydrodynamics of the disk that are
fundamental for determining its observational appearance
\citep{Blaes2014}.  It is also observationally difficult to isolate
the intrinsic disk continuum due to line-emission, host-galaxy
starlight, and internal reddening, and the emission peak is generally
unobservable due to absorption by intervening hydrogen.  After
accounting for these effects, it is sometimes possible to fit the
observed spectrum with disk models (e.g., \citealt{Capellupo2015}),
but this is not always the case (e.g., \citealt{Shankar2016}).
Alternative attempts to isolate the continuum emission have made use
of polarimetry \citep{Kishimoto2004, Kishimoto2008} and difference
spectra/color variability
\citep{Whilhite2005,Pereyra2006,Schmidt2012}, but the interpretation
of these data is not straightforward (e.g., \citealt{Kokubo2015,
  Kokubo2016}).

Reverberation mapping (RM, \citealt{Blandford1982, Peterson1993,
  Peterson2014}) is a powerful tool for the investigation of AGN
accretion disks.  The basic principle of RM is to search for temporal
correlations between the time-variable emission at different
wavelengths, which encode information about unresolved structures in
the AGN.  The formalism for RM is a convolution operation
\begin{align}
  f_R(t) = \int f_S(t-\tau)\Psi(\tau)\,d\tau,
\end{align}
where $f_S$ is the driving signal light curve, $f_R$ is the
reverberating light curve, $\tau$ is the time delay (or ``lag''), and
$\Psi(\tau)$ is the transfer function.  The transfer function is
determined by the matter distribution surrounding the source of $f_S$,
and the main goal of RM is to infer $\Psi(\tau)$ from observations of
$f_S$ and $f_R$.  Recovering the transfer function is an ill-posed
inverse problem that requires regularization \citep{Horne1994,
  Skielboe2015} or forward modeling \citep{Pancoast2014b,Starkey2016}
to solve.  However, one can still infer a great deal of information
from the cross-correlation of $f_S$ and $f_R$ to determine the mean
lag $\langle \tau \rangle$ between the light curves, which is related
to the first moment of $\Psi(\tau)$.  Combined with the speed of
light, the lag determines the characteristic size of the reverberating
structure.

In the context of the accretion disk, reverberation signals are
expected because of reprocessing arguments: shorter-wavelength
emission originates near the black hole where the disk is hottest,
while longer-wavelength emission originates in the cooler parts of the
disk at larger radii.  Self-irradiation by short-wavelength emission
deposits energy in the outer part of the disk, contributing an extra
heating term \citep{Shakura1973}.  As the short-wavelength emission
varies, it drives variations at longer wavelengths delayed by the
light travel time across the disk.  This model predicts that
short-wavelength variations will lead long-wavelength variations after
a time delay that scales with the size of the disk (e.g.,
\citealt{Krolik1991}).

Measuring these inter-band continuum lags is extremely difficult
because the predicted size of the accretion disk is only about one
light day (about 170 gravitational radii for a $10^8\,{\rm
M}_{\odot}$\ black hole), and monitoring campaigns require
comparable or better cadence to resolve such a short lag.  The first
suggestive (2--3$\sigma$) report was by \citet{Collier1998}, who found
that the UV variations led the optical variations in NGC\,7469.  Since
then there have been many hints of longer lags at longer wavelengths
in other AGN, but most of these measurements are not statistically
significant ($\leq\!  2\sigma$) and represent upper limits
\citep{Sergeev2005, Arevalo2008,Breedt2010, Lira2015, Troyer2016,
  Jiang2017,  Gliozzi2017, Buisson2017}.  The relationship between the
X-ray and optical emission also appears to be complex and does not
necessarily fit into a simple reprocessing model.  In some cases,
there is not enough energy in the X-ray variations to drive the long
term trends in the optical light curves \citep{Uttley2003,
  Arevalo2009, Breedt2009}.  There are also reports of optical
emission leading the X-rays \citep{Marshall2008}, as well as
uncorrelated X-ray/optical emission \citep{Maoz2002}.

The first secure detection ($>\!3\sigma$) of inter-band continuum lags
was in NGC\,2617 by \citet{Shappee2014}, who found longer lags at
longer wavelengths that were consistent with predictions for
reprocessing in a standard geometrically thin accretion disk.  The
only other AGN with significant continuum lag detections is NGC\,5548.
\citet{Mchardy2014} resolved continuum lags in this object using two
years of {\it Swift} data, while the Space Telescope and Optical
Reverberation Mapping project (AGN STORM, \citealt{DeRosa2015})
obtained the most complete RM measurement of the accretion disk to
date.  The STORM project detected inter-band lags between the X-ray,
UV, optical, and near IR wavelengths using four space-based
observatories and 25 ground-based telescopes (\citealt{Edelson2015,
  Fausnaugh2016, Starkey2017}), and the measured lag-wavelength
relation is again consistent with predictions for reprocessing in a
geometrically thin disk.  However, the size of the disk indicated by
the STORM measurements is larger by a factor of three than the
predictions from standard models.

This finding is consistent with results from gravitational
microlensing of strongly lensed quasars.  Microlensing is one of the
only other ways of investigating physical scales close to the SMBH,
and studies using this method also find disk sizes larger than
thin-disk theory by a factor of a few \citep{Morgan2010,
Blackburne2011, Mosquera2013, Jimenez-Vicente2014}.  However,
microlensing can only probe the disks in distant, high luminosity
quasars, while RM provides a means of probing accretion disks in
local, low-luminosity It is therefore imperative to
expand the sample of objects with secure inter-band continuum lags,
and several such programs have been completed, with others still in progress
(e.g., NGC\,4151, \citealt{Edelson2017}; NGC\,4593, \citealt{McHardy2016}).

In this study, we present detections of near-UV and optical inter-band
continuum lags in two Seyfert 1 galaxies, MCG+08-11-011 and NGC\,2617.
These objects were observed as part of a monitoring campaign in 2014,
the original goal being to measure SMBH masses using
continuum-H$\beta$ reverberations.  \citet{Fausnaugh2017b} presented
the spectroscopic monitoring component and the initial results for the
broad line lags and SMBH masses.  Here, we analyze four months of
densely sampled (0.6--1.5 day cadence) broad-band photometric
monitoring data for these objects and measure the inter-band continuum
lags.  MCG+08-11-011 is a new addition to the sample of objects with
secure accretion disk RM measurements.  NGC\,2617 is in a lower
luminosity state than when observed by \citet{Shappee2014}, and
provides an interesting reference point for investigating the
dependence of accretion disk structure on luminosity.
 
\floattable
\begin{deluxetable}{rccccc}
\tablewidth{0pt} \tablecaption{Physical Parameters\label{tab:targets}}
\tablehead{\colhead{Object} & \colhead{Redshift} & \colhead{Mass} &
  \colhead{Luminosity} & \colhead{Eddington Ratio} &
  \colhead{Accretion Rate}\\ &\colhead{$z$} &\colhead{$M_{\rm BH}$
    (M$_{\odot}$)} & \colhead{$L$ (erg s$^{-1}$)}&\colhead{$\dot
    m_{\rm Edd}$} &\colhead{(M$_{\odot}$
    yr$^{-1}$)}\\ \colhead{(1)}&\colhead{(2)}&\colhead{(3)}&\colhead{(4)}&\colhead{(5)}&\colhead{(6)}
} \startdata MCG+08-11-011 & 0.0205 & $(2.82 ^{+5.50}_{-1.86}) \times
10^{7} $ & $(1.98\pm 0.20)\times 10^{44} $ & 0.054 & $3.5\times
10^{-2} $ \\ NGC 2617 & 0.0142 & $(3.24 ^{+6.31}_{-2.14}) \times
10^{7} $ & $(4.27\pm 0.43)\times 10^{43} $ & 0.010 & $7.5\times
10^{-3} $ \\ \enddata \tablecomments{Columns 2, 3 and 4 are taken from
  \citet{Fausnaugh2017b}.  Column 3 was derived from
  continuum-H$\beta$ lags, and includes all systematic uncertainties.
  Column 4 is calculated from the observed mean optical luminosity,
  assuming a bolometric correction of 10 ($L = 10\lambda
  L_{5100\angstrom}$).  Note that this value has been corrected for
  Galactic extinction and host-galaxy starlight.  Columns 6 was
  calculated assuming a radiative efficiency of $\eta = 0.1$ (see \S2
  and Equation~\ref{equ:edd}).}
\end{deluxetable}

In \S2, we discuss our observations, data reduction, and light curves.
In \S3, we describe our time series analysis and present inter-band
continuum lags.  In \S4, we report results from a physical model of
our data using the Continuum REprocessed AGN Markov Chain Monte Carlo
code ({\tt CREAM}, \citealt{Starkey2016}).  In \S5, we discuss our
results in the context of an accretion-disk reprocessing model, and we
summarize our findings in \S6.  We assume a consensus cosmology with
$H_0 = 70 {\rm\ km\ s^{-1}\ Mpc^{-1}}$, $\Omega_{\rm m} = 0.3$, and
$\Omega_{\Lambda} = 0.7$.

\section{Targets and Observations}
The RM campaign extended between January and July of 2014, targeting
11 AGN, and was primarily based on observations at the MDM observatory
(see \citealt{Fausnaugh2017b} for details).  Supplemental data were
contributed by telescopes from around the globe, and a unique addition
over typical RM campaigns was broad-band imaging data in multiple
filters.  MCG+08-11-011 and NGC\,2617 were the two most variable AGN
during this campaign, and yielded robust measurements of the
continuum-H$\beta$ lags used to measure the SMBH masses
\citep{Fausnaugh2017b}.  Because of the strong variability signals, we
focused on these two objects for our first analysis of the broad-band
imaging data.

Table \ref{tab:targets} presents some of the important physical
parameters of these AGN.  The SMBH mass $M_{\rm BH}$ was determined by
\citet{Fausnaugh2017b} using 5100\,\AA\ continuum-H$\beta$ lags.  The
luminosity $L$ was derived from the average values of the 5100\,\AA\
continuum light curves of the same RM campaign and assuming a
bolometric correction of 10 (we discuss alternative bolometric
corrections in \S4 below).  Note that these estimates are corrected
for Galactic extinction and host-galaxy starlight (see
\citealt{Fausnaugh2017b} for details).  Defining the Eddington
luminosity and accretion rate as
\begin{align}
  L_{\rm Edd} = \frac{4\pi G M_{\rm BH} c}{\kappa}\label{equ:edd2}
\end{align}
and
\begin{align}
\dot M_{\rm Edd}
  = \frac{L_{\rm Edd}}{\eta c^2},
\label{equ:edd}
\end{align}
where $\kappa$ is the Thomson opacity ($\sim\! 0.4$\,cm$^2$\,g$^{-1}$)
and $\eta$ is the radiative efficiency, we calculate the Eddington
ratio $\dot m_{\rm Edd} = \dot M/\dot M_{\rm Edd} = L/L_{\rm Edd} $
and mass accretion rate $\dot M$ setting $\eta = 0.1$.

NGC\,2617 was observed with {\it Swift} during this time as part of a
continuing target-of-opportunity program (\citealt{Shappee2014}).
Daily exposures were taken with all six filters of the
UltraViolet/Optical Telescope (UVOT, \citealt{Roming2005}) with
simultaneous monitoring from the X-ray Telescope (XRT,
\citealt{Burrows2005}).

Our ground-based imaging is mostly from the {\it Las Cumbres
  Observatory} (LCO) 1m global telescope network \citep{Brown2013},
  acquired as part of the AGN Key project \citep{Valenti2015}.  The
  network consist of nine identical one-meter telescopes distributed
  at four sites around the world.  Each telescope has the same optical
  design and detectors.  At the time of the RM campaign, the detectors
  were SBIGSTX-16803 cameras with a field of view of $16' \times 16'$
  and a pixel scale of $0\farcs 23$.  Data were obtained between 2014
  January and 2014 May for MCG+08-11-011 and NGC\,2617 in the {\it
  ugriz} bands on an approximately daily cadence.  These filters have central
  wavelengths of 3600\,\AA, 4800\,\AA, 6300\,\AA, 7700\,\AA, and
  9100\,\AA, respectively (see Tables~\ref{tab:MCG+08-11-011_lags}
  and \ref{tab:NGC 2617_lags} for corrections to the rest-frame).
  MCG+08-11-011 has a high declination ($+46.5^{\circ}$) and can only
  be observed from LCO's northern-most site at McDonald Observatory.
  NGC\,2617 was observed from the sites at McDonald, Siding Spring,
  Australia, and Sutherland, South Africa.

Both objects were also observed with the 0.7m telescope at the Crimean
Astrophysical Observatory (CrAO).  Images were taken in the Johnson {\it
BVRI} bands with central wavelengths of 4400\,\AA, 5500\,\AA,
7000\,\AA, and 8800\,\AA, as well as a filter approximating the
Cousins {\it I}-band, designated {\it R1} which has a central wavelength
of 7900\,\AA, but is much narrower than the Johnson {\it I}-band.  The median
cadence of these observations is about 2 days.  The telescope is
equipped with a AP7p CCD detector that has a $15'
\times 15'$ field-of-view and pixel scale of $1\farcs 76$.  Finally,
{\it V}-band images were obtained with the 0.9m telescope at West
Mountain Observatory and the 0.5m telescope at Wise Observatory
\citep{Brosch2008}.

The spectroscopic monitoring observations were obtained at the MDM
observatory using the Boller \& Chivens CCD spectrograph on the 1.3m
Mcgraw-Hill telescope.  We extracted light curves for the rest-frame
5100\,\AA\, continuum, a region of the spectrum relatively free of
line emission.  The 2.3m telescope at Wyoming Infrared Observatory
(WIRO) contributed four epochs of optical spectroscopy to help fill
anticipated gaps in the MDM observations.  These data and our
inter-calibration procedures are described in detail by
\citet{Fausnaugh2017b}.

\subsection{Image Subtraction}
We analyzed the ground-based imaging data using the image subtraction
package {\tt ISIS} \citep{Alard1998}.  Although a common reduction
procedure was applied to all images, we analyzed the datasets from
each telescope and in each filter separately, including the individual
1m telescopes in the LCO network.

First, the raw images were bias-subtracted and flat-fielded at the
contributing facility using the appropriate software reduction
pipelines.  Next, all images were collected in a central repository
and vetted by eye for poor observing conditions or errors in the
initial image processing.  For each telescope, we then registered the
images to a common coordinate system, and we constructed a
high-quality reference image by combining the epochs with the best
seeing and lowest backgrounds.  Finally, we subtracted the reference
from each epoch using {\tt ISIS}.  {\tt ISIS} transforms the
point-spread-function (PSF) and flux scale of one image to match that
of a second image by fitting for a spatially variable convolution
kernel.  The subtraction leaves a clean measurement of the variable
flux on a pixel-by-pixel basis.

\subsection{Light curves} 
\begin{figure*}
\centering
\includegraphics[width=\textwidth]{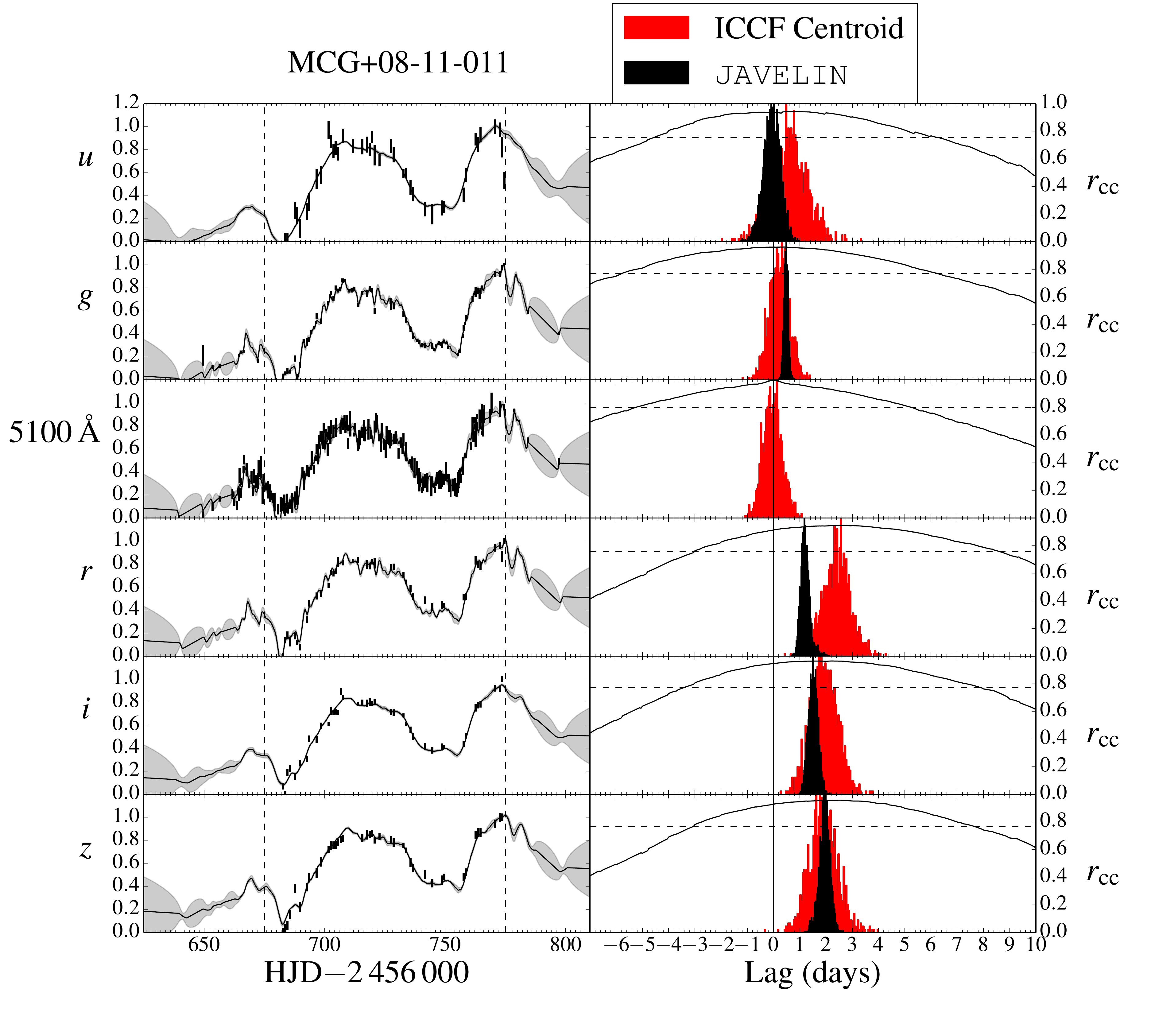}
\caption{Left panel: Light curves of MCG+08-11-011.  The {\it ugriz}
  data are from differential broad-band photometry (with 1$\sigma$
  uncertainties), while the 5100\,\AA\ light curve combines
  spectroscopy and \V\ imaging.  The y-axes are in flux units and
  scaled so that the minimum value of the light curve is 0 and the
  maximum value is 1.  Vertical dashed lines show the restricted
  temporal baseline used to calculate the interpolated
  cross-correlation function (ICCF).  The solid black lines and shaded
  regions show the {\tt JAVELIN} interpolation and their 1$\sigma$
  uncertainties.  Right panel: Lag estimates relative to the
  5100\,\AA\ continuum.  The black lines show the ICCF (or the
  autocorrelation function for the 5100\,\AA\ continuum).  The
  horizontal dashed lines show the threshold value of correlation
  coefficient $r_{\rm cc} = 0.8r_{\rm max}$ used to calculated the
  ICCF centroid.  The red histograms show the ICCF centroid
  distributions from the FR/RSS method (\S3), and the black histograms
  show the {\tt JAVELIN} posterior lag
  distributions.  \label{fig:mcg0811lags}}
\end{figure*}
\begin{figure*}
\centering
\includegraphics[width=\textwidth]{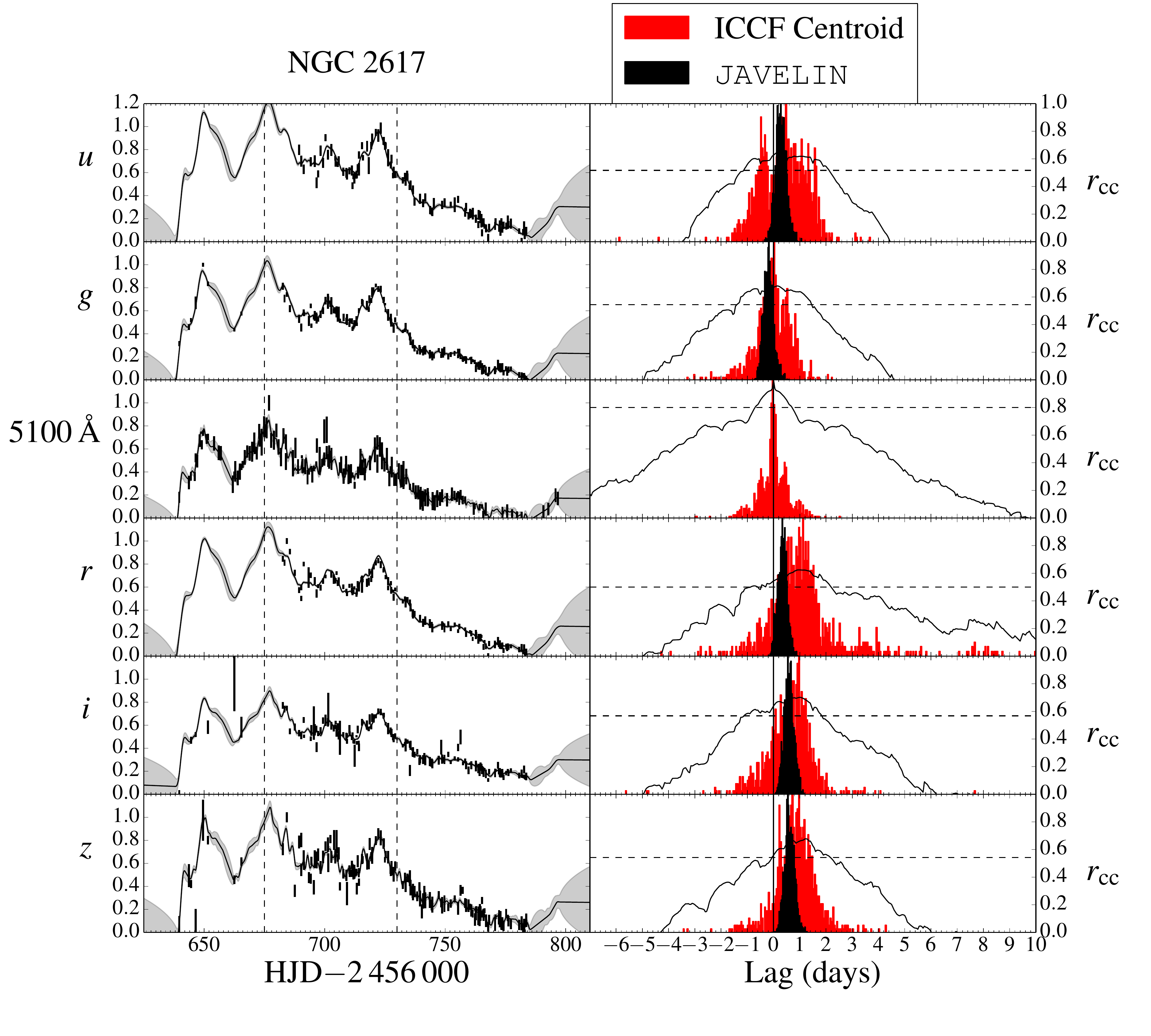}
\caption{Same as Figure \ref{fig:mcg0811lags}, but for the ground-based
  light curves of NGC 2617. \label{fig:n2617lags}}
\end{figure*}
\begin{figure*}
\centering
\includegraphics[width=\textwidth]{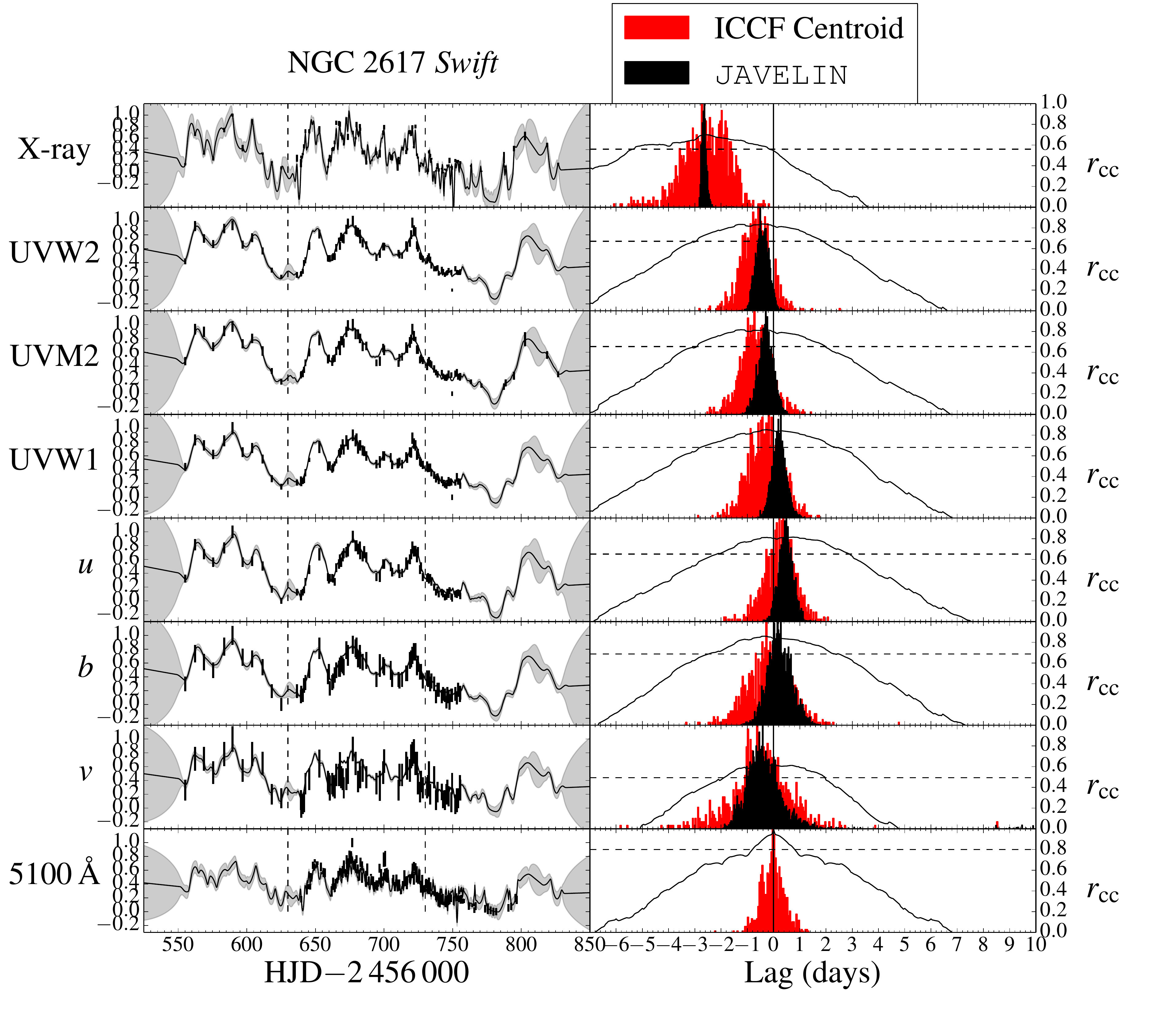}
\caption{Same as Figure \ref{fig:mcg0811lags}, but for the {\it Swift}
  light curves of NGC 2617. \label{fig:n2617_swiftlags}}
\end{figure*}

The {\it Swift} UVOT light curves were extracted using standard
aperture photometry techniques.  We used a 5\farcs0 radius circular
aperture centered on the AGN using the UVOT software task {\it
uvotsource}.  Background counts were estimated using the mode of pixel
values in a surrounding annulus 15\farcs0 in width.  The large
annulus was chosen to sample the background sky level, so this
procedure introduces a constant level of contamination from the
host-galaxy starlight within the 5\farcs0 aperture---however, the
contamination does not affect our final results, which only depend on
the differential variations of the light curves.  The XRT data were
reduced with the {\it xrtpipeline} task included in the HEASOFT
package, using the same apertures, response files, and modeling
techniques described in \citet{Shappee2014}.

For the ground-based data, we extracted differential light curves from
the subtracted images using the photometry package included with {\tt
  ISIS}.  First, the software fits a model to the reference image PSF.
Then, for all sources identified in the reference image, the software
smooths the model PSF by the convolution kernel fit during the image
subtraction and uses the result to perform PSF photometry on the
subtracted image.  The result is a light curve in units of
differential counts relative to the flux of the object in the
reference image.  These flux variations are free of constant
contaminates, such as host-galaxy starlight, and extrinsic variations
due to seeing or aperture effects.

{\tt ISIS} accounts for only the local Poisson uncertainty on the
observed counts.  To account for any systematic issues associated with
the image subtraction, we rescaled the light curve uncertainties to
match the residuals of comparison stars in the field-of-view.  Our
method closely follows that of \citet{Fausnaugh2016}.  For each epoch,
we compared the differential flux of each star to {\tt ISIS}'s
estimate of its uncertainty by calculating the rescaling factor
required to make the flux residuals consistent with zero at 1$\sigma$.
We then rescaled the flux uncertainties at that epoch by the median of
the rescaling factors of all comparison stars.  We imposed a minimum
rescaling factor of 1, since the photon noise sets a fundamental floor
on the precision, and we removed obvious variable stars by censoring
light curves with long term trends or mean rescaling factors greater
than 100.  On average, this forces the comparison star light curves to
have a reduced $\chi^2$ of 1 for a constant model.  We generally found
rescaling factors ranging between 1.0 and 6.0, depending on the
quality of the data and the number of comparison stars.  One LCO
telescope at Siding Spring had rescaling factors that reached 10 and
12 in the \usdss\ and \g, while one LCO telescope at Sutherland had
rescaling factors between 5 and 8 for all bands.  These telescopes had
fewer observations overall, limiting our ability to construct a good
reference image, and the image subtraction quality suffers as a
result.

To combine light curves from different telescopes, we used the
inter-calibration procedure described by \citet{Fausnaugh2016}.
Briefly, the calibration solves for maximum likelihood offsets and
rescaling factors, which account for the different flux levels in the
reference images of each telescope and the different definitions of
counts (due to heterogeneous detector responses, filter throughputs,
gains, etc).  To first order, the linear calibration model also
accounts for slight variations in the effective wavelengths and widths
of the different filter pass-bands.  Due to the limited amount of CrAO
data, we combined the \B\ with the \g, the \R\ with the \r, the {\it
  R1}-band with the \isdss, and the \I\ with the \z.  The CrAO \V\
light curve was incorporated with the spectroscopic continuum light
curve from MDM.

Because observations at different telescopes are never taken
simultaneously, it is necessary to interpolate the light curves when
solving for the calibration shifts and rescaling factors.  We
predicted the light curve values at intermediate times using  {\tt
  JAVELIN}  \citep*{Zu2011}, which models the light curves
with a stochastic process model.  Process models with different
covariances/power spectra are available with {\tt JAVELIN}, but we
have found that the damped-random walk (DRW, or Ornstein-Uhlenbeck
process) is adequate for this purpose.  As discussed in
\citet{Fausnaugh2017b}, our light curves are not long enough to
constrain the damping time scale of the process model, so we fixed
this parameter to 200 days (see also \citealt{Kozlowski2017}).

Finally, we flux-calibrated the differential light curves by
performing aperture photometry on the reference image of one
standardized data set.  We chose the McDonald LCO data as the
standard, since this light curve has the largest number of
observations.  All other light curves are transformed to match the
flux scale and mean value of this light curve using the {\tt JAVELIN}
intercalibration routine.  Flux calibration then reduces to measuring
the reference image's zeropoint magnitude and the total counts of the
AGN.  We used a 5\farcs0 radius circular aperture and a sky annulus of
15\farcs0.  We did the same for all comparison stars, and measured AB
magnitude zeropoints in each image by matching to the SDSS DR7
photometric catalog \citep{Abazajian2009}.  The final flux
measurements are again contaminated by the host-galaxy starlight in
the reference image, though this contamination does not contribute to
the variations measured from the image subtraction.  See
\citet{Fausnaugh2016} for a more
thorough discussion of this flux-calibration technique.

The light curves are given in Tables~\ref{tab:mcg0811lc} through
\ref{tab:n2617swiftlc}, and shown in the left-hand panels of
Figures\,\ref{fig:mcg0811lags}--\ref{fig:n2617_swiftlags}.
Table~\ref{tab:lc_prop} summarizes useful properties of the light
curves.  For MCG+08-11-011, the median cadence is about 1 day (1.5
days in the \usdss), and for NGC\,2617, the median cadence is about
0.6 days in the {\it ugriz} bands and 1.1--1.2 days for the {\it
  Swift} data.  We also self-consistently estimated the mean flux
$\hat F$ and the intrinsic variability $\sigma_{\rm var}$ (corrected
for measurement noise) by solving for the values of these parameters
that minimize
\begin{align}
  -2 \ln \mathcal{L} = \sum_i^{N_t} \frac{\left[ F(t_i) - \hat
      F\right]^2}{\sigma^2(t_i) +  \sigma_{\rm var}^{2} } + \sum_i^{N_t} \ln
  \left[\sigma^2(t_i) + \sigma_{\rm var}^{ 2} \right],
\label{equ:fracvar}
\end{align}
where $F(t_i)$ is the flux measurement at time $t_i$ and
$\sigma_(t_i)$ is its uncertainty.  This procedure is identical to
that of \citet{Fausnaugh2017b}, and we similarly report the rms
fractional variability amplitude $F_{\rm var} = \sigma_{\rm var}/\hat
F$, the mean signal-to-noise $\langle S/N \rangle$, and the
signal-to-noise of $F_{\rm var}$
\begin{align} {\rm (S/N)_{var}} = \frac{\sigma_{\rm var}}{\bar \sigma
    \sqrt{ 2/N_{\rm obs}}}\label{equ:sn_var}
\end{align}
for each light curve, where $\bar \sigma$ is the mean measurement
uncertainty among observations and $N_{\rm obs}$ is the number of
observations.  The right hand side of
Equation \ref{equ:sn_var} is derived by assuming that the variance of
$\bar \sigma$ has a reduced $\chi^2$ distribution.

\begin{deluxetable}{lrcr}
\tablecaption{MCG+08-11-011 Light Curves\label{tab:mcg0811lc}}
\tablehead{\colhead{Filter} & \colhead{HJD $-$2\,400\,000} & \colhead{$F_{\lambda}$} & \colhead{Telescope ID}\\
& \colhead{(days)} & \colhead{($10^{-15}$erg\,cm$^{-2}$s$^{-1}$\AA$^{-1}$)}
}
\startdata
{\it u} & 56683.5826 & $4.9925 \pm 0.1199$ & LCO1 \\
{\it u} & 56684.5841 & $4.9352 \pm 0.1691$ & LCO1 \\
{\it u} & 56687.5629 & $5.5462 \pm 0.1651$ & LCO1 \\
\dots&\dots&\dots&\dots\\
\hline
{\it g} & 56639.5200 & $6.4602 \pm 0.0266$ & CrAO \\
{\it g} & 56649.4759 & $6.8878 \pm 0.1825$ & CrAO \\
{\it g} & 56653.4950 & $6.6852 \pm 0.0309$ & CrAO \\
\dots&\dots&\dots&\dots\\
\hline
{\it r} & 56682.5935 & $8.1803 \pm 0.0420$ & LCO1 \\
{\it r} & 56683.5838 & $8.4089 \pm 0.0454$ & LCO1 \\
{\it r} & 56684.5854 & $8.4846 \pm 0.0546$ & LCO1 \\
\dots&\dots&\dots&\dots\\
\hline
{\it i} & 56682.5940 & $6.4003 \pm 0.0316$ & LCO1 \\
{\it i} & 56683.5992 & $6.3211 \pm 0.0344$ & LCO1 \\
{\it i} & 56684.5858 & $6.5392 \pm 0.0365$ & LCO1 \\
\dots&\dots&\dots&\dots\\
\hline
{\it z} & 56682.5945 & $5.7181 \pm 0.0297$ & LCO1 \\
{\it z} & 56683.5997 & $5.7453 \pm 0.0306$ & LCO1 \\
{\it z} & 56684.5863 & $5.7661 \pm 0.0318$ & LCO1 \\
\dots&\dots&\dots&\dots\\
\enddata
\tablecomments{A machine-readable version of this table is published in the
  electronic edition of this article. A portion is shown here for
  guidance regarding its form and content.}
\end{deluxetable}

\begin{deluxetable}{lrcr}
\tablecaption{NGC~2617 Light Curves\label{tab:n2617lc}}
\tablehead{\colhead{Filter} & \colhead{HJD $-$2\,400\,000} & \colhead{$F_{\lambda}$} & \colhead{Telescope ID}\\
& \colhead{(days)} & \colhead{($10^{-15}$erg\,cm$^{-2}$s$^{-1}$\AA$^{-1}$)}
}
\startdata
{\it u} & 56689.3963 & $6.9971 \pm 0.0770$ & LCO5 \\
{\it u} & 56690.2849 & $6.8172 \pm 0.0719$ & LCO5 \\
{\it u} & 56690.2916 & $6.8395 \pm 0.0701$ & LCO4 \\
\dots&\dots&\dots&\dots\\
\hline
{\it g} & 56639.6731 & $6.3221 \pm 0.2172$ & CrAO \\
{\it g} & 56643.6272 & $7.6720 \pm 0.2321$ & CrAO \\
{\it g} & 56644.5132 & $8.2198 \pm 0.1750$ & CrAO \\
\dots&\dots&\dots&\dots\\
\hline
{\it r} & 56682.6001 & $9.9075 \pm 0.0397$ & LCO5 \\
{\it r} & 56683.3380 & $9.8560 \pm 0.0415$ & LCO4 \\
{\it r} & 56684.3056 & $10.1500 \pm 0.0392$ & LCO6 \\
\dots&\dots&\dots&\dots\\
\hline
{\it i} & 56639.6672 & $6.9188 \pm 0.0114$ & CrAO \\
{\it i} & 56644.5162 & $7.0678 \pm 0.0096$ & CrAO \\
{\it i} & 56646.5303 & $7.0834 \pm 0.0067$ & CrAO \\
\dots&\dots&\dots&\dots\\
\hline
{\it z} & 56639.6682 & $6.1075 \pm 0.0665$ & CrAO \\
{\it z} & 56643.6256 & $6.4154 \pm 0.0692$ & CrAO \\
{\it z} & 56644.5170 & $6.3643 \pm 0.0238$ & CrAO \\
\dots&\dots&\dots&\dots\\
\enddata
\tablecomments{A machine-readable version of this table is published in the
  electronic edition of this article. A portion is shown here for
  guidance regarding its form and content.}
\end{deluxetable}

\begin{deluxetable}{lcc}
\tablecaption{NGC~2617 {\it Swift} Light Curves\label{tab:n2617swiftlc}}
\tablehead{\colhead{Filter} & \colhead{HJD $-$2\,400\,000} & \colhead{$F_{\lambda}$} \\
& \colhead{(days)} & \colhead{($10^{-15}$erg\,cm$^{-2}$s$^{-1}$\AA$^{-1}$)}
}
\startdata
{X-rays\tablenotemark{a}} & 56413.9240 & $2.8400\pm 0.1200$ \\
{X-rays} & 56415.2573 & $2.2900\pm 0.1200$ \\
{X-rays} & 56415.6885 & $2.3400\pm 0.1200$ \\
\dots& \dots &\dots \\
\hline
{\it UVW2} & 56413.9240 & $12.1448 \pm 0.5593$ \\
{\it UVW2} & 56415.2573 & $12.3706 \pm 0.6836$ \\
{\it UVW2} & 56415.6885 & $11.9232 \pm 0.6589$ \\
\dots& \dots &\dots \\
\hline
{\it UVM2} & 56413.9240 & $11.9514 \pm 0.6605$ \\
{\it UVM2} & 56415.2573 & $12.2862 \pm 0.6790$ \\
{\it UVM2} & 56415.6885 & $12.0620 \pm 0.6666$ \\
\dots& \dots &\dots \\
\hline
{\it UVW1} & 56413.9240 & $10.6184 \pm 0.5868$ \\
{\it UVW1} & 56415.2573 & $10.9159 \pm 0.6032$ \\
{\it UVW1} & 56415.6885 & $10.5211 \pm 0.5814$ \\
\dots& \dots &\dots \\
\hline
{\it u} & 56413.9240 & $8.8754 \pm 0.4087$ \\
{\it u} & 56415.2573 & $8.6335 \pm 0.3976$ \\
{\it u} & 56415.6885 & $8.4759 \pm 0.3903$ \\
\dots& \dots &\dots \\
\hline
{\it b} & 56413.9240 & $6.9168 \pm 0.3185$ \\
{\it b} & 56415.2573 & $6.6667 \pm 0.3070$ \\
{\it b} & 56415.6885 & $6.7283 \pm 0.3099$ \\
\dots& \dots &\dots \\
\hline
{\it v} & 56413.9240 & $6.5059 \pm 0.2397$ \\
{\it v} & 56415.2573 & $6.6268 \pm 0.2441$ \\
{\it v} & 56415.6885 & $6.5059 \pm 0.2397$ \\
\dots& \dots &\dots \\
\enddata
\tablenotetext{a}{{\raggedright 0.3--10 keV absorption-corrected flux ($10^{-11}$erg\,cm$^{-2}$s$^{-1}$).\\}}
\tablecomments{A machine-readable version of this table is published in the
  electronic edition of this article. A portion is shown here for
  guidance regarding its form and content.}
\end{deluxetable}

\floattable
\begin{deluxetable}{llrrcrrrr}
\tablewidth{0pt}
\tablecaption{Light-curve Properties \label{tab:lc_prop}}
\tablehead{\colhead{Object} & \colhead{Light curve} & \colhead{N$_{\rm obs}$} & \colhead{$\Delta t_{\rm med}$} & \colhead{$\hat F$} & \colhead{$\langle {\rm S/N} \rangle$} & \colhead{$F_{\rm var}$} & \colhead{${\rm (S/N)_{var}}$}\\
& & &\colhead{(days)}  & & &\\
\colhead{(1)}&\colhead{(2)} &\colhead{(3)}&\colhead{(4)}&\colhead{(5)}&\colhead{(6)}&\colhead{(7)}&\colhead{(8)}
}
xs\startdata
MCG+08-11-011 & {\it u} & 41 & 1.54 & 6.30 & 44.3 & 0.10 & 19.0 \\
 & {\it g} & 85 & 0.99 & 7.47 & 202.2 & 0.07 & 83.3 \\
 & {\it r} & 42 & 1.07 & 9.17 & 194.4 & 0.05 & 41.0 \\
 & {\it i} & 41 & 1.07 & 7.01 & 231.7 & 0.04 & 41.1 \\
 & {\it z} & 41 & 1.07 & 6.23 & 225.5 & 0.03 & 33.5 \\
\hline
NGC\,2617 & X-rays & 136 & 1.20 & 4.00 & 20.8 & 0.57 & 91.2 \\
 & UVW2 & 126 & 1.12 & 13.50 & 18.9 & 0.40 & 59.2 \\
 & UVM2 & 126 & 1.27 & 12.50 & 17.6 & 0.36 & 50.3 \\
 & UVW1 & 131 & 1.09 & 11.00 & 18.8 & 0.29 & 42.9 \\
 & {\it Swift u} & 130 & 1.10 & 8.89 & 21.5 & 0.29 & 49.2 \\
 & {\it Swift b} & 129 & 1.11 & 6.94 & 21.6 & 0.16 & 26.9 \\
 & {\it Swift v} & 119 & 1.14 & 6.43 & 22.2 & 0.09 & 15.3 \\
 & {\it u} & 113 & 0.66 & 7.11 & 77.5 & 0.10 & 56.6 \\
 & {\it g} & 166 & 0.56 & 8.38 & 275.1 & 0.04 & 83.8 \\
 & {\it r} & 127 & 0.62 & 10.60 & 319.2 & 0.04 & 82.4 \\
 & {\it i} & 154 & 0.60 & 8.28 & 287.7 & 0.02 & 46.9 \\
 & {\it z} & 153 & 0.59 & 8.42 & 230.7 & 0.02 & 41.6 \\
\enddata
\tablecomments{Column 3 gives the number of observations in each light
  curve. Column 4 gives the median cadence.  Column 5 gives the mean
  flux level of each light curve in units of $10^{-15}$
  erg\,cm$^{-2}$\,s$^{-1}$\,\AA$^{-1}$ (the X-rays are units of
  $10^{-11}$erg\,cm$^{-2}$\,s$^{-1}$).  Column 6 gives the mean
  signal-to-noise ratio $\langle {\rm S/N} \rangle$.  Column 7 gives
  the rms fractional variability, defined in \S2.2.  Column 8 gives
  the approximate S/N at which we detect variability (\S2.2).  }
\end{deluxetable}

\section{Time Series Analysis}

We searched for lags in the continuum light curves using the
interpolated cross correlation function (ICCF) and the Bayesian model
of {\tt JAVELIN}.  Full descriptions of the cross-correlation
technique can be found in \citet{Gaskell1987}, \citet{White1994}, and
\citet{Peterson2004}, while a complete description of {\tt JAVELIN}
can be found in \citet{Zu2011, Zu2013}.

In brief, the ICCF method uses piecewise linear interpolation to
estimate the cross-correlation coefficient $r_{\rm cc}$ for two light
curves after shifting one by a given lag $\tau$.  We evaluated the
ICCF on a grid of lags spaced by 0.05 days, and we estimated the lag
between the two light curves using the ICCF centroid.  The ICCF
centroid $\tau_{\rm cent}$ is defined as the average $\tau$ weighted
by $r_{\rm cc}$ for $r_{cc} > 0.8r_{\rm max}$, where $r_{\rm max}$ is
the maximum of the ICCF.  For completeness, we also report the lag
$\tau_{\rm peak}$ that corresponds to $r_{\rm max}$.  Uncertainties on
the lag were estimated using the flux randomization/random subset
sampling (FR/RSS) method of \citet{Peterson2004}.  Individual points
from each light curve were resampled (with replacement), adjusted by
random Gaussian deviates scaled to the measurement uncertainties, and
the centroid $\tau_{\rm cent}$ was recalculated.  After repeating this
procedure $10^3$ times, the central 68\% interval of the resulting
centroid distribution was adopted for the uncertainty in $\tau_{\rm
  cent}$.

{\tt JAVELIN} determines the lags between light curves by modeling the
data as a stochastic process (a DRW) and fitting for the transfer
function $\Psi(\tau)$ (see \S1).  The formalism assumes that the
transfer function can be approximated by a top-hat, parameterized by a
scaling factor, width, and central time delay.  The central time delay
is adopted as a measure of the lag, which we designate $\tau_{\tt
  JAV}$.  It has been shown by \citet{Skielboe2015} that measurements
of the lag do not depend on the choice of stochastic process used to
describe light curve variations.  {\tt JAVELIN} can also fit multiple
light curves and their underlying lags simultaneously, which maximizes
the amount of information used in the fit and accounts for covariances
between the lags of different light curves (note that there is no
prior on the relations between the transfer functions for the
different light curves).

In each case, we measured the lags relative to the 5100\,\AA\ light
curve.  This choice is unimportant for {\tt JAVELIN}, but in the ICCF analysis it is important to use the best light curve in terms of sampling and noise properties as the reference light curve.  For the ICCF method, we shifted and
interpolated both light curves, and used the average value of $r_{\rm
cc}$ to estimate $\tau_{\rm cent}$.  We restricted the light curve
baselines to 6675 $<$ HJD$-$2\,450\,000$<$ 6775 days for
MCG+08-11-011, to avoid interpolating over large gaps in the \g.  For
NGC 2617, we restricted the baseline to 6675 $<$ HJD$-$2\,450\,000 $<$
6730 days, to avoid the gradual flux variations at the tail of the
light curves.  Gradual variations such as these can affect the ICCFs
due to red-noise leakage (\citealt{Welsh1999}, see
also \citealt{Fausnaugh2017b}).  The {\it Swift} light curves begin
somewhat earlier than the ground-based data (6630 days), and we
include these earlier observations in our analysis.  We did not
otherwise detrend the data.

For the {\tt JAVELIN} models of MCG+08-11-011, we fit all of the {\it
  ugriz} data simultaneously.  For NGC\,2617, the combination of {\it
  Swift} and {\it ugriz} light curves was too large for {\tt JAVELIN}
to converge on a solution in a reasonable amount of time, so we fit
the {\it Swift} and ground-based data sets separately.  {\tt JAVELIN}
removes any linear trends from the light curves in the fits, and we
did not limit the temporal baselines when fitting with {\tt JAVELIN}.

\subsection{Results}

In Figures\,\ref{fig:mcg0811lags}--\ref{fig:n2617_swiftlags} we show
the ICCFs, lag centroid distributions, and {\tt JAVELIN} posterior lag
distributions for each light curve.  Table
\ref{tab:MCG+08-11-011_lags} gives the lags and their uncertainties
for MCG+08-11-011, and Table \ref{tab:NGC 2617_lags} gives the same
for NGC\,2617, both corrected to the rest-frame.  The lags derived
from the ICCF and {\tt JAVELIN} approaches are consistent, except for
the \r\ in MCG+08-11-011: the ICCF centroid distribution gives a
rest-frame lag of $2.56\pm 0.51$ days, while {\tt JAVELIN} finds a lag
of $1.19 \pm 0.16$ days.  We found that this difference is related to
interpolation of the \r\ light curve over the large gaps in the second
half of the campaign.  If we only interpolate the 5100\,\AA\ continuum
light curve, the ICCF lag is $\tau_{\rm cent} = 2.19 \pm 0.59$ days,
which reduces the discrepancy from 2.7$\sigma$ to 1.7$\sigma$.  The
widths of the {\tt JAVELIN} posteriors are much smaller than the ICCF
centroid distributions, and for NGC\,2617, we must rely on {\tt
  JAVELIN} to claim statistically significant detections.  The {\tt
  JAVELIN} distributions are also narrower for the {\it Swift} data,
although the UV--optical lags are only detected at the
0.9--1.5$\sigma$ level, which may be related to the longer cadence of these light curves.  The uncertainty in the lag from the ICCF
method is intrinsically limited by the width of the autocorrelation of
the continuum light curve \citep{Peterson1993}, and, given the better
precision using {\tt JAVELIN}, we adopt $\tau_{\tt JAV}$ for our final
lag measurements.  {\tt JAVELIN} also accounts for the correlations
between lags from light curves at different wavelengths, and therefore
maximizes the amount of information used in the fit.  Overall, the
uncertainties on the lag are unlikely to be any smaller than the
estimates from {\tt JAVELIN}, while the ICCF centroid distributions
probably place upper limits on the lag uncertainties.

Our results are largely consistent with a disk reprocessing model,
with larger lags at longer wavelengths.  In fact, the trends are
nearly monotonic, with the main exceptions being the \usdss\ lags in NGC\,2617 and the \g\ lag in MCG+08-11-011.  The
{\it Swift} UVW1, {\it b}, and {\it v} band lags in NGC\,2617 are also
contrary to this trend, but are consistent with 0 days at less than
1$\sigma$.

The \usdss\ lags in NGC 2617 are detected at
1.5$\sigma$ and 2.0$\sigma$, respectively.  These filters are
contaminated by Balmer continuum emission from the BLR, which is
expected to reverberate on longer time scales than the continuum
emission and may bias the observed lags to larger values.  This bias
has been seen in NGC\,5548 \citep{Edelson2015, Fausnaugh2016} and the
{\it Swift} monitoring data of NGC\,2617 from 2013 analyzed by
\citet{Shappee2014}.

The \g\ lag relative to 5100\,\AA\ in MCG+08-11-011 of $0.50 \pm 0.08$
days is detected at high significance (6.25$\sigma$).  It is less
clear what might be affecting this band, so we investigated the cause
of the lag in more detail.  One possibility is that the uncertainties
on the data are underestimated.  To check this, we re-ran the FR/RSS
procedure and the {\tt JAVELIN} fits with the \g\ uncertainties
inflated by factors of 1.5 and 3.0 since the 5100\,\AA\ light curve
uncertainties are unlikely to be underestimated (see the detailed
explanation in \citealt{Fausnaugh2017}).  For the ICCF centroid
distributions, the median lag did not change, although the width of
the distributions increased.  For the {\tt JAVELIN} models, the
posterior lag distributions shifted closer to zero lag---for the
$\times$1.5 rescaling, the lag is $0.36\pm 0.20$
days (1.8$\sigma$), and for the $\times$3.0 rescaling, the
lag is $0.22\pm 0.25$ days
(0.9$\sigma$).

This seems to indicate that the \g\ lag is an artifact.  However, it
is peculiar that the ICCF centroid, which relies on different
assumptions than {\tt JAVELIN} and is less dependent on the
measurement uncertainties, should consistently be skewed away from
zero.  Further investigation showed that the positive \g-5100\,\AA\
lag signal is weakly present in both the LCO and CrAO light curves
independently, with a lag of $0.33\pm 0.35$ days for LCO and $0.31 \pm
0.15$ days for CrAO (rest frame).  We also tried fitting the {\it
V}-band data (which tends to have smaller uncertainties) separately
from the MDM spectroscopic 5100\,\AA\ light curve.  This still yielded
a positive \g\ lag, with values of $0.71\pm 0.12$ days
(5.9$\sigma$) relative to the {\it V}-band and $0.42 \pm 0.15$ days
(2.8$\sigma$) relative to the 5100\,\AA\ continuum.

Inspection of Table \ref{tab:MCG+08-11-011_lags} and Figure
\ref{fig:mcg0811lags} show that the lag-wavelength relation through
the {\it ugriz} bands is monotonic.  Thus, another possibility is that
the 5100\,\AA\ light curve is an outlier and that these results are
related to using this light curve as the driver.  To test this, we
re-ran the FR/RSS procedure and the {\tt JAVELIN} models using the \g\
as the driving light curve.  However, this made no change except to
shift all of the observed lags by precisely the \g--5100\,\AA\ lag.

Thus, there is some evidence that the lag is a real signal in the
data.  A possible explanation is bias by BLR emission, similar to the
\usdss\ and {\it Swift} \usdss\ in NGC\,2617.  The \g\ is contaminated
by both the H$\beta$ and $H\gamma$ broad lines in MCG+08-11-011, while
the spectroscopic data and {\it V}-band are virtually free of line
emission (there may be a small amount of BLR contamination by Fe{\sc
  ii} emission at these wavelengths). Using the {\tt synphot} package
in {\tt IRAF} to estimate broad-band fluxes from the mean MDM
spectrum, we find that the Balmer lines contribute only 7\% of the
total \g\ continuum flux, so it would be surprising if line emission
had a large effect on the observed lag.  However,
\citet{Fausnaugh2016} found that the bias from BLR emission depends
more strongly on the variability amplitude of the line emission, which
is quite large in this object (the Balmer lines display fractional
variability amplitudes $F_{\rm var} > 7$--9\%,
\citealt{Fausnaugh2017b}).  Thus, it is not out of the question that
the 0.2 to 0.5 day lag is biased by BLR emission.  If a similar bias
exists in the other broad-band filters, this may explain why the
5100\,\AA\ light curve appears as an outlier from the lag-wavelength
relation.  The \usdss\ and \r\ are contaminated by Balmer continuum
emission and H$\alpha$, respectively, while the Paschen continuum may
be significant in the \isdss\ and \z\ \citep{Korista2001}.

\floattable
\begin{deluxetable}{rrrrrrr}
\tablewidth{0pt}
\tablecaption{MCG+08-11-011 Rest-frame Continuum Lags \label{tab:MCG+08-11-011_lags}}
\tablehead{
\colhead{Filter} & \colhead{$\lambda$} & \colhead{$\tau_{\rm cent}$} & \colhead{$\tau_{\rm peak}$} & \colhead{$\tau_{\tt JAV}$}  & \colhead{$\tau_{\rm CREAM} - \tau_{5100\angstrom}$}& \colhead{$\tau_{\tt JAV} - \tau_g$} \\
&\colhead{(\AA)}&\colhead{(days)}&\colhead{(days)}&\colhead{(days)}&\colhead{(days)}\\
\colhead{(1)}&\colhead{(2)}&\colhead{(3)}&\colhead{(4)}&\colhead{(5)}&\colhead{(6)}&\colhead{(7)}
}
\startdata
{\it u} & 3449 & $0.66_{-0.60}^{+0.68}$ & $-0.05_{-1.30}^{+0.86}$   & $-0.04^{+0.30}_{-0.32}$ & $-0.52\pm 0.16$  & $-0.54^{+0.30}_{-0.32}$    \\
{\it g} & 4703 & $0.78_{-0.36}^{+0.30}$ & $0.38_{-0.53}^{+0.43}$    & $0.50^{+0.08}_{-0.07}$   & $-0.18\pm 0.24$ & $0.00^{+0.08}_{-0.07}$      \\
5100\,\AA & 5100 & $-0.01_{-0.31}^{+0.30}$ & $0.00_{-0.14}^{+0.14}$ & \dots               & $0.00\pm 0.27$     &   $-0.50$                 \\
{\it r} & 6124 & $2.49_{-0.51}^{+0.49}$ & $2.16_{-0.48}^{+0.72}$    & $1.19^{+0.16}_{-0.15}$  & $0.24\pm 0.33$   & $0.69^{+0.16}_{-0.15}$       \\
{\it i} & 7535 & $2.02_{-0.55}^{+0.49}$ & $1.92_{-0.58}^{+0.91}$    & $1.52^{+0.20}_{-0.18}$  & $0.70\pm 0.43$   & $1.02^{+0.20}_{-0.18}$     \\
{\it z} & 8927 & $1.94_{-0.57}^{+0.48}$ & $1.97_{-0.58}^{+0.82}$    & $1.94^{+0.19}_{-0.20}$  & $1.20\pm 0.55$   & $1.44^{+0.19}_{-0.20}$    \\
\enddata
\tablecomments{Column 2 gives the rest-frame effective wavelength of
  the filter.  Column 3 gives the ICCF centroids and the 68\%
  confidence intervals from the FR/RSS procedure (see \S3).  Column 4
  gives the same but for the ICCF peaks. Column 5 gives the lags fit
  by {\tt JAVELIN} and the central 68\% confidence interval of the
  posterior distributions.  Column 6 gives the lag estimates from the
  transfer functions fit by {\tt CREAM} (see \S4).  Column 7 is the
  same as Column 5 with the \g\ lag subtracted.  All lag values have
  been corrected to the rest-frame.}
\end{deluxetable}

\floattable
\begin{deluxetable}{rrrrrrrrr}
\tablewidth{0pt}
\tablecaption{NGC 2617 Rest-frame Continuum Lags \label{tab:NGC 2617_lags}}
\tablehead{\colhead{Filter} & \colhead{$\lambda$} & \colhead{$\tau_{\rm cent}$} & \colhead{$\tau_{\rm peak}$} & \colhead{$\tau_{\tt JAV}$} & \colhead{$\tau_{\rm CREAM}-\tau_{5100\angstrom}$} \\
&\colhead{(\AA)}&\colhead{(days)}&\colhead{(days)}&\colhead{(days)}&\colhead{(days)}\\
\colhead{(1)}&\colhead{(2)}&\colhead{(3)}&\colhead{(4)}&\colhead{(5)}&\colhead{(6)}
}
\startdata
X-rays & 9 & $-2.47_{-0.81}^{+0.88}$ & $-2.48_{-1.17}^{+1.22}$ & $-2.58^{+0.09}_{-0.08}$ & \dots   \\
UVW2 & 1900 & $-0.63_{-0.48}^{+0.51}$ & $-0.58_{-0.63}^{+0.78}$ & $-0.39^{+0.25}_{-0.25}$ & $-0.38\pm 0.03$ \\
UVM2 & 2214 & $-0.66_{-0.55}^{+0.52}$ & $-0.92_{-0.97}^{+0.49}$ & $-0.25^{+0.28}_{-0.28}$ & $-0.35\pm 0.03$ \\
UVW1 & 2563 & $-0.29_{-0.64}^{+0.62}$ & $-0.24_{-0.68}^{+0.83}$ & $0.24^{+0.27}_{-0.25}$ & $-0.33\pm 0.04$  \\
{\it Swift u} & 3451 & $0.27_{-0.51}^{+0.56}$ & $0.19_{-0.92}^{+0.68}$ & $0.49^{+0.26}_{-0.25}$ & $-0.24\pm 0.06$ \\
{\it u} & 3470 & $0.40_{-0.86}^{+0.89}$ & $0.24_{-1.17}^{+0.83}$ & $0.31^{+0.20}_{-0.19}$ & $-0.24\pm 0.06$  \\
{\it Swift b} & 4268 & $-0.12_{-0.74}^{+0.78}$ & $-0.24_{-1.02}^{+0.87}$ & $0.23^{+0.47}_{-0.40}$ & $-0.01\pm0.08$  \\
{\it g} & 4732 & $0.03_{-0.58}^{+0.58}$ & $0.10_{-0.63}^{+0.92}$ & $-0.16^{+0.18}_{-0.18}$ & $-0.08\pm 0.09$ \\
5100\,\AA & 5100 & $0.02_{-0.47}^{+0.47}$ & $0.00_{-0.19}^{+0.19}$ & \dots & $0.00\pm 0.11$ \\
{\it Swift v} & 5326 & $-0.46_{-1.02}^{+0.96}$ & $-0.29_{-1.12}^{+1.22}$ & $-0.38^{+0.64}_{-0.52}$ & $0.01\pm 0.11$ \\
{\it r} & 6162 & $0.98_{-1.01}^{+0.91}$ & $0.92_{-0.87}^{+1.51}$ & $0.37^{+0.17}_{-0.18}$ & $0.11\pm 0.13$ \\
{\it i} & 7582 & $0.68_{-0.58}^{+0.82}$ & $0.63_{-0.78}^{+0.92}$ & $0.60^{+0.20}_{-0.18}$ & $0.32\pm 0.17$ \\
{\it z} & 8982 & $0.86_{-0.61}^{+0.64}$ & $0.78_{-0.68}^{+0.63}$ & $0.62^{+0.20}_{-0.18}$ & $0.54\pm 0.21$  \\
\enddata
\tablecomments{Columns 2 through 6 are the same as in
  Table\,\ref{tab:MCG+08-11-011_lags}.  All values have been corrected
  to the rest-frame.  }
\end{deluxetable}

\section{{\tt CREAM} Modeling}

\begin{figure}
\centering
\includegraphics[width=0.5\textwidth]{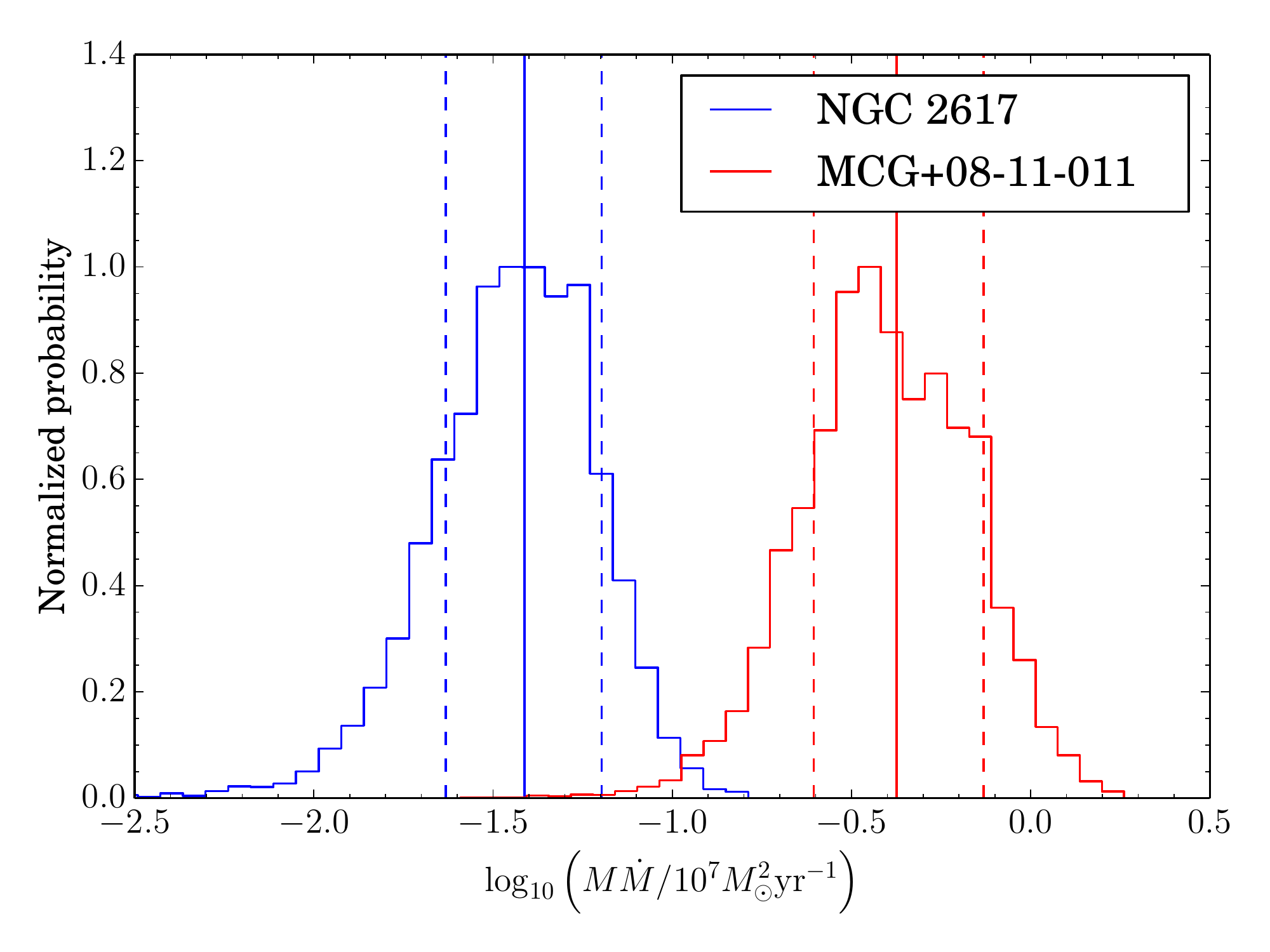}
\caption{Posterior distributions for $M_{\rm BH}\dot M$ for
  MCG+08-11-011 (red) and NGC 2617 (blue), as derived by {\tt CREAM}.
  The solid lines give the medians of the distributions, the dashed
  lines give the 68\% confidence intervals.  For these fits, the
  temperature profile was fixed to $R^{-3/4}$ and the inclination to
  $i = 0$. \label{fig:n2617_mmdot}}
\end{figure}

We also analyzed the light curves using the Continuum REprocessed AGN
Markov Chain Monte Carlo code ({\tt CREAM}, \citealt{Starkey2016}).
{\tt CREAM} fits an accretion disk reprocessing model directly to the
observed light curves so as to estimate the temperature profile of the
disk and its inclination to the observer's line-of-sight.  The adopted
geometry is a ``lamp post,'' which attributes the primary emission to
a point source at a small distance above the central black hole.  As
the lamp post varies, it irradiates the disk, which thermally
reprocesses the incident flux into variations at longer wavelengths.
Physically, the point source may correspond to the X-ray emitting
corona, if the corona is small compared to the size of the disk.
However, a physical interpretation of the lamp post is not
required---this geometry should reasonably approximate most models
that place the origin of driving emission near the SMBH at a small height above the disk midplane.  
This includes models like those of \citet{Luo1998} and \citet{Nealon2015}, which modify the structure of the disk on scales of $\sim 10R_{\rm g}$.

{\tt CREAM} fits the model by inferring the transfer functions and
driving lamp-post light curve that best reproduces the observed data.
The transfer functions are calculated from a thin accretion disk with
three parameters: the temperature $T_0$ at the inner-edge, the index
$\beta$ of a power law temperature profile, and the inclination $i$ of
the disk to the observer's line of sight.  {\tt CREAM} takes a
Bayesian approach, sampling the posterior probability distributions of
the driving light curve and disk parameters.  With an estimate of
$T_0$, it is then possible to calculate the product $M_{\rm BH}\dot
M$, where $M_{\rm BH}$ is the mass of the black hole and $\dot M$ is
the mass accretion rate through the disk (see \citealt{Cackett2007}
and \citealt{Starkey2016} for a derivation).

For our first model, we fixed the power law index $\beta$ to $-3/4$
and held the inclination $i$ of the disk constant at 0 degrees.  In
Figure\,\ref{fig:n2617_mmdot}, we show the posterior distributions of
$M_{\rm BH}\dot M$ for the two AGN (we show fits to the transfer
functions of individual light curves and the inferred driving light
curves in the Appendix).  {\tt CREAM} finds $\log M_{\rm BH}\dot M =
6.63 \pm 0.24$\, $[M^2_\odot$\,yr$^{-1}]$\ in MCG+08-11-011, and $\log
M_{\rm BH}\dot M = 5.58 \pm 0.21$\,$[M^2_\odot$\,yr$^{-1}]$ in NGC
2617.  We then ran models that allowed the inclination to vary (owing
to the short wavelength baseline spanned by our light curves, we were
unable to place meaningful constraints on $\beta$).  For
MCG+08-11-011, we were unable to constrain the inclination, but for
NGC\,2617 we found $i = 43\pm 20$ degrees with $\log M_{\rm BH}\dot M
= 5.24 \pm 0.23$\,$[M^2_\odot$\,yr$^{-1}]$.

We compare the {\tt CREAM} results with our time-series analysis from
\S3 by giving the mean lags of the transfer functions in
Tables\,\ref{tab:MCG+08-11-011_lags} and \ref{tab:NGC 2617_lags}.
Since we use the 5100\,\AA\ light curve as a reference in \S3, we
subtract this lag from the other {\tt CREAM} values in these tables.
For NGC\,2617, there is excellent agreement between the {\tt CREAM}
results and the lags estimated from $\tau_{\rm cent}$ and $\tau_{\tt
  JAV}$.  For MCG+08-11-011, the {\tt CREAM} lags are shifted by about
0.5 to 1.0 days relative to the values of $\tau_{\tt JAV}$.  As noted
in \S3.1, the \g-5100\,\AA\ lag from our time series analysis may be
an outlier, while the temperature gradient in {\tt CREAM}'s physical
model forces $\tau \propto \lambda^{4/3}$ and an anomalous
\g-5100\,\AA\ lag cannot be produced. Using the \g\ as the reference
wavelength (subtracting the \g\ lag from the other lags in Columns
3--6 of Table\,\ref{tab:MCG+08-11-011_lags}), we find much better
agreement.  This calculation is explicitly shown for $\tau_{\tt JAV}$
in Table\,\ref{tab:MCG+08-11-011_lags}.  The reduced $\chi^2$ values
of the {\tt CREAM} fits are larger than would be expected for Gaussian
statistics (1.96--2.37), which may indicate that the light curve
uncertainties are underestimated, or that the model is not a perfect
description of the data.

\citet{Fausnaugh2017b} estimated black hole masses for these objects,
which allows us to calculate $\dot M$ from the {\tt CREAM} fits
(Table~\ref{tab:targets}).  For MCG+08-11-011, $M_{\rm BH}
\sim\!2.82\times 10^{7} $\,M$_{\odot}$, implying $\dot M =
0.151$\,M$_\odot$\,yr$^{-1}$ and an Eddington ratio $\dot m_{\rm Edd}
= \dot M/\dot M_{\rm Edd} = 0.234$ with $\eta = 0.1$.  For NGC 2617,
$M_{\rm BH} \sim\!  3.24\times 10^{7}$\,M$_{\odot}$ and we calculate
$\dot M = 0.012$\,M$_\odot$\,yr$^{-1}$\ and $\dot m_{\rm Edd} =
0.016$.

These Eddington ratios can be compared to independent estimates using
the observed luminosities during the monitoring campaign
(Table\,\ref{tab:targets}, again, we assume that $\dot M/\dot M_{\rm
Edd} = L/L_{\rm Edd}$, $L = 10\lambda L_{5100\angstrom}$).  The
estimates of the Eddington ratios from the {\tt CREAM} models are a
factor of 4.3 larger for MCG+08-11-011 and a factor 1.6 larger for
NGC\,2617.  However, there are large uncertainties associated
with these estimates.  \citet{Runnoe2012} empirically find a bolometric correction of
$8.1\pm 0.4$, but recommend estimating the bolometric luminosity with
the relation $L \approx 10^{4.9}(\lambda L_{\rm
5100\angstrom})^{0.9}$, with an intrinsic scatter of 0.17 dex around
this relation.  The first option would decrease the value of $\dot
m_{\rm Edd}$ (as estimated from the optical luminosity) by about 20\%,
while the second estimate would increase $\dot m_{\rm Edd}$ by 20\%
and 38\% in MCG+08-11-011 and NGC\,2617, respectively.  Thus, there is
at least a factor of 1.2 to 1.3 systematic uncertainty on $\dot m_{\rm Edd}$ as
estimated from the optical luminosity, with an additional factor of
1.5 statistical uncertainty due to the intrinsic scatter.\footnote{With the multi-wavelength coverage for NGC~2617, we also estimated the bolometric correction by integrating the observed mean fluxes corrected for Galactic extinction and then dividing by the mean of $L_{5100}$.  This yields a bolometric correction of 9.3.  Alternatively, we fit the composite QSO template of \citet{VandenBerk2001} to the UV data, and integrated the X-rays and template through 1$\mu$m.  This yields a slightly smaller bolometric correction of 8.9, since this method is not affected by  host-galaxy light.}  The widths of the 68\% confidence
intervals of the posterior distributions of $M \dot M$ also imply a
factor of $\sim$1.7 uncertainty on the estimate of $\dot m_{\rm Edd}$
from {\tt CREAM}. Furthermore, the estimate of $\dot m_{\rm Edd}$ from
{\tt CREAM} depends on the adopted SMBH mass, which is intrinsically
uncertain by a factor of 2.5 to 3.0.  Finally, the estimate of $\dot
m_{\rm Edd}$ from {\tt CREAM} depends on the choice of $\eta$.
Although $\eta$ can vary between about $0.06$ and $ 0.50$, depending
on the spin of the black hole, it is most likely that the black hole
is co-rotating with the disk.  This suggests that $\eta \ge 0.1$,
which decreases $\dot M_{\rm Edd}$ and increases the discrepancy in
$\dot m_{\rm Edd}$.  We therefore ignore uncertainty in $\eta$ when
calculating the significance of the discrepancies, noting that
uncertainty in this parameter will tend to strengthen our results.

Combining in quadrature the uncertainties on the bolometric
correction, $M\dot M$ from {\tt CREAM}, and $M_{\rm BH}$, there is a
total uncertainty of about 0.56 dex.  The factor of 4.3 disagreement
in MCG+08-11-011 can then be written as $0.64\pm 0.56$ dex, while the
factor of 1.6 disagreement in NGC\,2617 is $0.19\pm 0.56$ dex.  The
estimate of $\dot m_{\rm Edd}$ from {\tt CREAM} for NGC\,2617
therefore appears to be consistent with the observed optical
luminosity, and there is only a small difference 
for MCG+08-11-011.

\section{Discussion}

\subsection{Comparison with Previous Studies}
MCG+08-11-011 is one of the two objects with statistically significant
continuum lags based on data from CrAO taken between 2001--2003, with
an approximately 3-day cadence excepting seasonal gaps
\citep{Sergeev2005}.  Based on the centroid of the cross correlation
functions (using the same ICCF and FR/RSS methods employed in this
study), \citet{Sergeev2005} found lags for the \V, \R, and \I\
(relative to the \B) of $0.91\pm0.53$ days, $4.64 \pm 0.81$, and
$5.75\pm 1.18$ days (rest-frame), respectively.  This result accords
with the expected trend of larger lags at longer wavelengths.
However, the magnitude of the lags is several days larger than those
measured here, which range between about 1 to 1.5 days from the \g\ to
the \isdss\ and \z.  While it is possible that the structure of the
disk has changed over the past decade, it is more likely that the
lower cadence and large gaps in the \citet{Sergeev2005} light curves
result in larger uncertainties than estimated in that study.  The
light curves presented here have no seasonal gaps and daily cadence,
which should yield more reliable lags.

Inter-band continuum lags were detected in NGC\,2617 by
\citet{Shappee2014}.  After the All Sky Automated Survey for
Supernovae
(ASAS-SN\footnote{\url{http://www.astronomy.ohio-state.edu/~assassin}})
observed a sudden X-ray/optical outburst of this AGN in 2013,
intensive multi-wavelength monitoring of the target ensued for $\sim\!
50$ days.  This target-of-opportunity campaign was led by X-ray and
near UV observations from the {\it Swift} satellite, while
ground-based monitoring extended the wavelength coverage through
optical and near-IR wavelengths.  The observed lags, relative to the
\V, ranged from $-1.11\pm 0.32$ days in the UVW2 filter to $1.97 \pm
1.32$ days in the \I\ and $7.42\pm 1.25$ days in the {\it K}-band
(rest-frame).  There was also a $3.36\pm 0.42$ X-ray lag (or a
$\sim\!2.2$ day lag between the X-rays and UVW2).  These lags were
measured with {\tt JAVELIN} in the same way as in this study, but
using the {\it Swift} X-ray and UVW2 light curves as drivers.

The UVW2--5100\,\AA\ lag measured here is almost a factor of 3 smaller
than that measured by \citet{Shappee2014}, although this is only a
2.3$\sigma$ difference. We find that the \isdss\ and \z\ lags from the
2014 data are also smaller than in 2013 by a factor of about 2
(1.0$\sigma$). The X-ray lag is 0.78 days (1.9$\sigma$) shorter than
that reported by \citet{Shappee2014}.

Thus, the lags measured here are broadly consistent with those
reported by \citet{Shappee2014}, but systematically smaller.  Under
the standard RM formalism, the observed lag between two light curves
is not independent of the driving light curve autocorrelation function
\citep{Blandford1982,Peterson1993}.  The variability amplitude of
NGC\,2617 in 2013 was much stronger than in 2014, and the time scale
of variations is smaller in 2014 than in 2013.  Both of these factors
will generally lead to smaller lags, which may account for these
results.  This is similar to the results of \citet{Goad2014}, who find
that lags in the BLR will be observed to be smaller for weak and rapid
variations simply due to geometric dilution.  Another possibility is
that the physical configuration of the disk has changed---because NGC
2617 is a ``changing look'' AGN, the accretion flow may be far from
equilibrium \citep{LaMassa2015, MacLeod2016, Runnoe2016}. The two
monitoring programs are separated by 1 year, and the dynamical time at
a distance of 1 light day from the black hole is about 1 month.  A
bulk readjustment of the accretion flow is therefore possible in the
time between the two campaigns.  The luminosity was also a factor of
1.8 smaller in 2014 compared to 2013, and the size of the disk is
expected to scale with luminosity (see \S5.3 below).  However, this
adjustment should happen on a viscous time scale, which is of order
decades to centuries for a typical Seyfert~1 (e.g.,
\citealt{LaMassa2015}).

\subsection{Challenges to the Disk Reprocessing Model}
The disk reprocessing model posits that short-wavelength radiation
drives long-wavelength emission by heating the accretion disk and
perturbing the local temperature.  Two important predictions of this
model are that the X-ray, UV, and optical light curves will be
well-correlated, and that longer wavelength light curves should lag
behind shorter wavelength light curves.  We qualitatively find results
consistent with disk reprocessing---the UV and optical light curves in
both objects are well correlated, and the lag-wavelength relation is
nearly monotonic.

However, it is clear from visual inspection of the NGC\,2617 light
curves that there is much more structure in the X-rays than in the UV
and optical emission, especially on short time scales.  Although the X-ray light curve would be expected to be smoothed if reprocessed at UV wavelengths, comparison of Figure\,\ref{fig:n2617_swiftlags} and Figure \ref{fig:n2617_cream1} shows that the inferred driving light curve does not correlate very well with the observed X-ray variations (this is confirmed by the ICCF analysis from \S3).    This poor correlation was also seen in 2013 by
\citet{Shappee2014}, and has been observed in other objects, including
NGC 5548 \citep{Uttley2003, Edelson2015}, MR 2251-178
\citep{Arevalo2008}, Mrk 79 \citep{Breedt2009}, NGC 3783
\citep{Arevalo2009}, and NGC 4151 (\citealt{Edelson2017}).  Several
of these studies have been unable to represent the UV/optical light
curves as a reprocessed (smoothed and shifted) version of the X-ray
light curve (\citealt{Arevalo2008,Breedt2009,Starkey2017}), which is
problematic for a generic disk reprocessing model.  A notable
exception is \citet{Shappee2014}, who were able to produce a good, but
not perfect, match between the X-ray and optical light curves from
2013 using a simple reprocessing model for NGC\,2617.  However, they
found X-ray to UV/optical lags (2--3 days) that are much larger than
any plausible light-travel time across the accretion disk.
\citet{Shappee2014} were unable to provide a physical interpretation
for the X-ray--optical lag, and we find a similar X-ray--optical lag
here ($\sim$2.6 days), reaffirming this problem for the disk
reprocessing model.

Although the X-ray light curve in NGC 2617 is problematic for disk
reprocessing, this paradigm may still be important---the UV/optical
light curves display strong correlations and follow the prediction of
larger lags at longer wavelengths.  A possible explanation is that the
driving light curve is in the extreme UV \citep{Shakura1973,
  Gardner2017}.  On the other hand, although the X-ray light curve has
additional structure compared to the UV/optical light curves, there is
still clearly some connection.  Considering the temporal lead of the
high-energy emission, this suggests a very complicated relationship
between the X-rays and UV/optical emission.  We stress that there are many possibilities for the geometry and energetics of the X-ray emitting corona based on both analytic results and simulations, and it is not clear what lag-wavelength relations or variable X-ray emission these models would produce (see, for example, \citealt{Schnittman2013a} and \citealt{Jiang2014b} for coronal emission extended across the disk and \citealt{Begelman2015, Sadowski2016, Begelman2017} for the possible effects of toroidal magnetic fields).  However, observational evidence strongly favors a compact corona, \citep{Reis2013, Mosquera2013}, which makes some aspects of the reprocessing model very likely,  and continuing
multi-wavelength monitoring of this and other Seyfert 1s is therefore
an important avenue for further investigations.  For example, see \citet{Giustini2017} for an analysis
of the X-ray emission in NGC 2617 during 2013, and
\citet{Oknyansky2017} for an analysis of the X-ray through IR emission
in 2016.

\subsection{Disk Radii and Temperature Profiles}
\begin{figure*}
\centering
\includegraphics[width=0.9\textwidth]{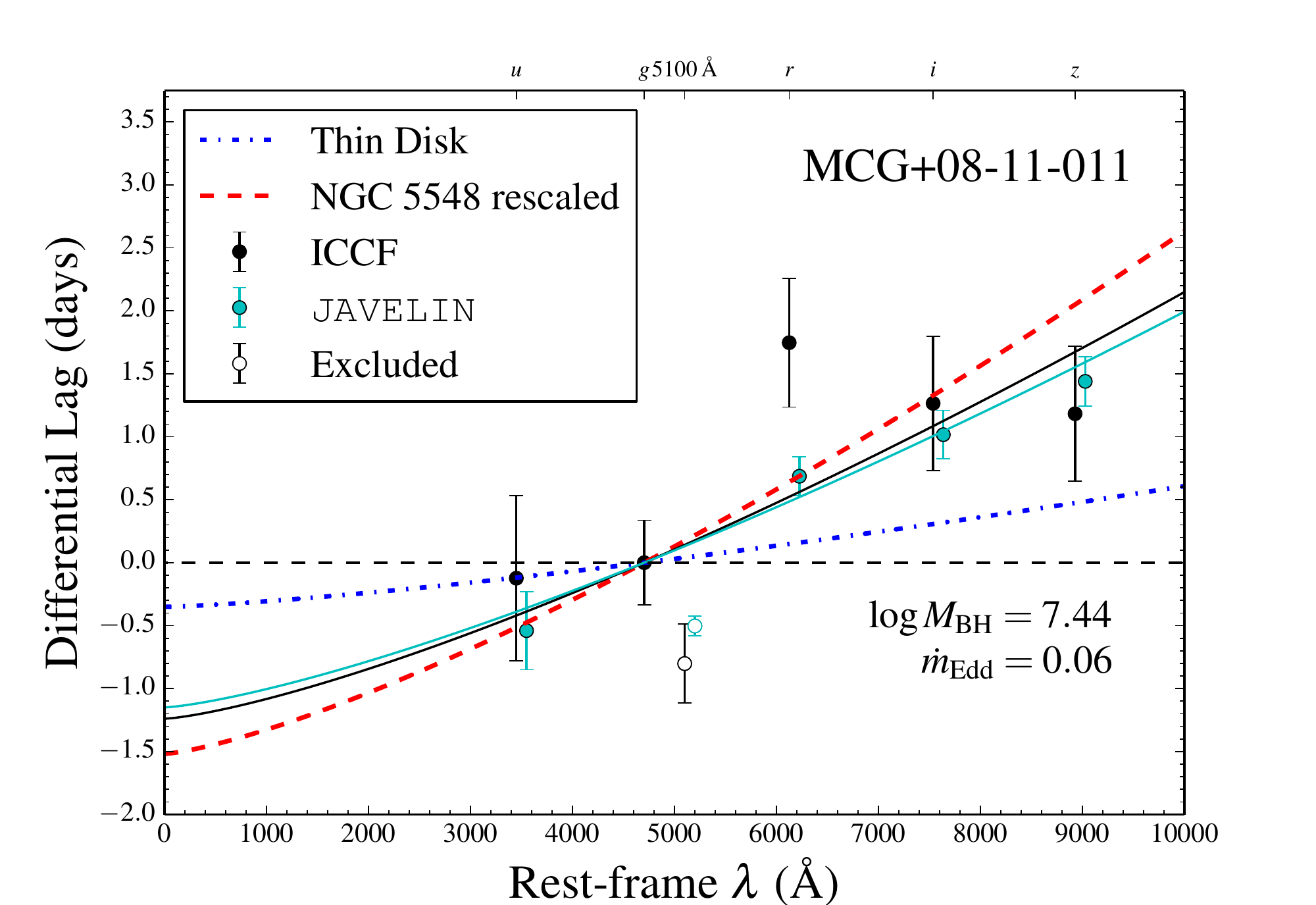}
\includegraphics[width=0.9\textwidth]{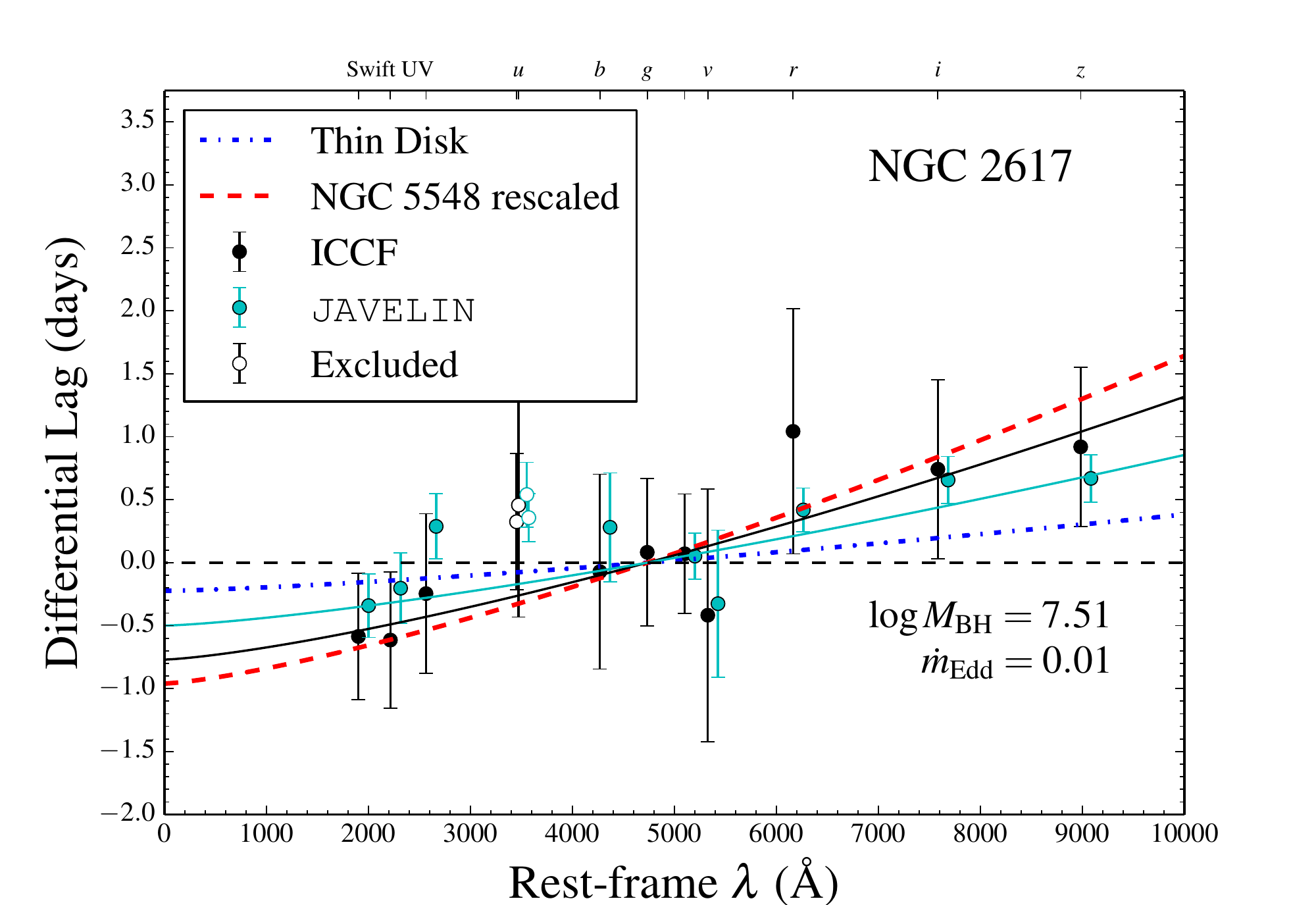}
\caption{Lag-wavelength relations for each object relative to the \g.
  The lags from the ICCF are shown in black and those from {\tt
    JAVELIN} are shown in cyan.  The best fit of $\tau_0$ with $\beta
  = 4/3$ are shown by the black and cyan lines (see \S5.3).  The
  predictions from standard thin-disk theory (Equation
  \ref{equ:tau_0_scale}) are shown by the dot-dashed blue lines, while
  the accretion disk in NGC\,5548 \citep{Fausnaugh2016}, rescaled to
  the mass and mass accretion rate of these objects, is shown by the
  dashed red lines.  \label{fig:disks}}
\end{figure*}

Figure \ref{fig:disks} shows the lags as a function of wavelength for
each object.  For a disk reprocessing model, the lag-wavelength
relation contains information about the absolute size of the disk and
the temperature profile.  To quantify this, we fit a model of the
form
\begin{align}
  \tau = \tau_0\left[ \left(\frac{\lambda}{\lambda_0}\right)^{\beta}
    -1 \right],
  \label{equ:lagmodel}
\end{align}
where $\lambda$ is the effective wavelength transformed to the
rest-frame, $\lambda_0$ is some reference wavelength, and $\tau_0$ and
$\beta$ are free parameters.  The normalization $\tau_0$ measures the
radius of the disk emitting at a reference wavelength $\lambda_0$, and
the index $\beta$ measures the temperature profile of the disk,
$T\propto R^{\, -1/\beta}$.  Standard thin-disk theory predicts that
$\beta = 4/3$, and assuming that the lags trace the flux-weighted mean
radius for emission at $\lambda$, $\tau_0$ scales as
\begin{align}
\left(\frac{\tau_0}{1.0 \,\,{\rm days}}\right) = 
\left(\frac{\lambda_0}{4800\,{\rm \angstrom} } \right)^{4/3}
 \left(\frac{\,M_{\rm BH}}{10^{8}{\rm M_{\odot}} }\right)^{2/3}
  \left(\frac{\dot m_{\rm Edd}}{0.09} \right)^{1/3} 
\label{equ:tau_0_scale}
\end{align}
\citep{Fausnaugh2016}.  This calculation assumes a radiative
efficiency $\eta = L_{\rm Bol}/\dot M c^2 = 0.10$ and that the
X-ray/far-UV radiation does not appreciably heat the disk compared to
viscous dissipation.  There can be deviations from this thin disk model on small scales ($\sim 10R_{\rm g}$ or a  few light hours, e.g., \citealt{Luo1998, Noble2011, Schnittman2016}), which would only result in a small modification to the lag-wavelength relation on the scales observed here (several light days).

We fit models with both $\tau_0$ and $\beta$ free to vary, as well as
with $\beta$ fixed to $4/3$.  We set the reference wavelength to
$\lambda_0 = 4800/(1+z)$\,\AA, the rest-frame \g\ effective
wavelength.  To match the model so that the \g\ lag equals 0, we
subtracted the \g--5100\,\AA\ lag from all measurements of $\tau_{\tt
  JAV}$ (as discussed in \S3.3, this is equivalent to fitting the lags
using the \g\ as the driver). We also tested fits where we excluded
suspect lag measurements.  In particular, we tried omitting the
\usdss\ lag for both targets because this lag is probably contaminated
by Balmer continuum emission from the BLR.  For MCG+08-11-011, we also
tried omitting the anomalous \g--5100\,\AA\ lag, which is likely an
outlier.  For NGC\,2617, we tested models that jointly fit the {\it
  Swift} and ground-based data, as well as fits to each dataset
separately.  We also excluded the large X-ray lag for this object,
since the lag is much larger than any plausible light-travel
time.

The results of these fits are given in Table\,\ref{tab:fits}.  Because
of the large uncertainties and limited amount of data, the fits
sometimes prefer a flat relation ($\beta = 0$), which does not provide
any constraint on $\tau_0$.  For the fits that do converge, the
uncertainties on $\tau_0$ and $\beta$ are still very large and do not
put interesting physical constraints on the disk.  We include the
results of these fits for completeness, but the rest of our discussion
focuses on the fits for $\tau_0$ assuming $\beta = 4/3$.

Fits to the ICCF and {\tt JAVELIN} lags give consistent values of
$\tau_0$, although the ICCF fits are poorly constrained.  Excluding
the anomalous \g--5100\,\AA\ lag in MCG+08-11-011 results in
reasonable values of $\chi^2/{\rm dof}$ (where ${\rm dof}$ is the
number of degrees-of-freedom in the fit) between 0.75 and 1.01 with
${\rm dof}$ between 3 and 4.  For NGC\,2617, including the \usdss\ and
{\it Swift} \usdss\ lags give $\chi^2/{\rm dof}$ of 1.67--2.66 with
${\rm dof}$ between 5 and 11, somewhat larger than would be expected
for Gaussian statistics.  Censoring these lags results in a
$\chi^2/{\rm dof}$ between 0.79 and 1.16.  Excluding all of the {\it
  Swift} data increases $\tau_0$ from 0.38 days to 0.51 days, while
excluding the ground-based data decreases $\tau_0$ to below 0.19 light
days.  This can be understood by the very small UV lags that are only
detected at $\sim\!  1\sigma$.  Including these data moves the model
to smaller $\tau_0$, consistent with the unresolved lags, while the
well-resolved ground-based lags pull $\tau_0$ to larger values.

The fits with $\chi^2/{\rm dof} \sim 1$ indicate that a disk
reprocessing model with $\beta = 4/3$ can reproduce our data very
well.  This is consistent with the prediction for a geometrically thin
disk with a temperature profile $T\propto R^{-3/4}$.  This signature
power law is difficult to reproduce if the disk is not geometrically
thin, and it is not immediately clear what other configurations could
mimic the $\tau \propto \lambda^{4/3}$ relation.  For MCG+08-11-011,
an acceptable $\chi^2/{\rm dof}$ requires the removal of the
\g--5100\,\AA\ lag, while for NGC 2617 we must exclude both the
ground-based and {\it Swift} {\it u}-bands.  As discussed above, we
already suspect that these lags are unreliable, so we adopt final
measurements for the disk sizes of $\tau_0 = 1.15\pm 0.11$  days
in MCG+08-11-011 and $\tau_0 = 0.50\pm 0.12$ days in NGC 2617.
These uncertainties represent only the formal errors in the fit.

Using the black hole masses and accretion rates in
Table\,\ref{tab:targets}, we can predict $\tau_0$ using
Equation\,\ref{equ:tau_0_scale}.  These values are given in
Table\,\ref{tab:fits} alongside our fits, and the predicted
lag-wavelength relations are shown by the dot-dashed blue lines in
Figure\,\ref{fig:disks}.  We find that our fits for $\tau_0$ are much
larger than these predictions.  For MCG+08-11-011, the disk is a
factor of 3.3 larger (a $7.2\sigma$ result), while for NGC\,2617 the
disk is a factor of 2.3 larger (a 2.3$\sigma$ result).

It is unclear if uncertainties in $M_{\rm BH}$ and $\dot m_{\rm Edd}$
can explain these discrepancies.  A factor of 3.3 increase in $\tau_0$
for MCG+08-11-011 requires a factor of 36 increase in the product
$M_{\rm BH}^2\dot m_{\rm Edd}$, and a factor of 12 increase for a
factor of 2.3 in $\tau_0$ in the case of NGC\,2617.  Even if the
values of $M_{\rm BH}$ from \citet{Fausnaugh2017b} are underestimated
by a factor of 3 (approximately equivalent to the intrinsic scatter in
the mean virial factor $\langle f \rangle$, \citealt{Onken2004}), the
optical luminosity would also have to underestimated $\dot m_{\rm
  Edd}$ by a factor of 1.4--4.0.  As discussed in \S4, the choice of
bolometric correction may be responsible for part of this difference.
Internal extinction, kinematic luminosity (i.e., energy-loss in
outflows), and advection may also be important, since these effects will cause the observed
luminosity to underestimate the true energy generation rate and the
inferred value of $\dot m_{\rm Edd}$.  Finally, the normalization of
Equation~\ref{equ:tau_0_scale} depends on the radiative efficiency
$\eta$ and relative heating by irradiation from X-rays/far UV
emission.  The radiative efficiency must be less than 0.1 to increase
the predicted size of $\tau_0$, which implies a counter-rotating black
hole relative to the disk and is \textit{a priori} unlikely, while
setting the heating term from irradiation to match that of viscous
dissipation only increases $\tau_0$ by $\sim$10\%
\citep{Fausnaugh2016}.  Given these uncertainties, there seems to be
no problem accounting for the discrepancy in NGC\,2617 ($\sim$40\%).
The factor of 4 discrepancy in MCG+08-11-011 is more difficult to
account for, although a combination of effects may be able to explain
the difference.

An even larger fluctuation (greater than a factor of 3) of $M_{\rm
  BH}$ beyond the estimate from \citet{Fausnaugh2017b} might cause the
discrepancies in disk size and $\dot m_{\rm Edd}$ to vanish.
Conversely, fluctuations of $M_{\rm BH}$ below these estimates would
cause severe energy budget problems, by about an order of magnitude
compared to the optical luminosity.  Uncertainty in the individual
estimates of $M_{\rm BH}$ therefore limit our ability to assess the
significance of any discrepancies with the thin disk model, and
improving these uncertainties is a critical path forward.

These results are consistent with our findings in \S4, where we showed
that the accretion rates inferred from {\tt CREAM} are formally larger
than those estimated from the optical luminosity.  As a reminder, {\tt
CREAM} finds discrepancies in $\dot m_{\rm Edd}$ of a factor of 4.3
and 1.3 in MCG+08-11-011 and NGC~2617, respectively, although these estimates are statistically consistent with 1 given the large uncertainties.  {\tt CREAM}
employs a physical model of the disk, which probably results in more
accurate estimates of the accretion rate than using the flux-weighted
mean radius in Equation~\ref{equ:tau_0_scale}.  This seems to suggest
that any discrepancy in disk size or $\dot m_{\rm Edd}$ is less severe
than indicated above.  However, the values of $\chi^2/{\rm dof}$ for
the {\tt CREAM} fits were somewhat larger than unity, which may
indicate that the {\tt CREAM} model does not adequately describe the
data.  Furthermore, the {\tt CREAM} uncertainties are large because
they again depend on the black hole mass.

If the disks are larger than expected from thin-disk theory, these
results are similar to those from RM of the accretion disk in NGC
5548.  \citet{Fausnaugh2016} find a disk in this object three times
larger than the prediction of standard thin-disk theory.  However,
they assumed that the accretion rate was 10\% of Eddington and that
irradiation by high-energy emission contributed significant heating to
the disk.  Based on optical spectroscopy taken during the AGN STORM
campaign \citep{Pei2017}, we measure the Eddington ratio of NGC 5548 in
2014 to be 5\% (again taking $\dot m_{\rm Edd} = 10\lambda L_{\rm
  5100\angstrom}/L_{\rm Edd}$).  Assuming that the X-rays/far UV
contribute negligible heating (as we did in
Equation~\ref{equ:tau_0_scale}), the disk in this object is a factor
of 4.4 larger than thin-disk theory.  As a comparison, we rescaled the
large disk from NGC 5548 to the mass and mass accretion rate of our
targets using the same dependencies as in
Equation\,\ref{equ:tau_0_scale}.  These comparisons are given in
Table\,\ref{tab:fits} and shown in Figure\,\ref{fig:disks} by the
dashed red lines.  Our fits lie in between the NGC 5548 result and the
prediction from thin-disk theory.  Unlike the case of NGC 5548, we are
forced to extrapolate the lag-wavelength relation to far UV
wavelengths.  Therefore, there are considerably larger uncertainties
associated with our estimate of the disk's absolute size.  However,
the qualitative agreement (an accretion disk larger than predictions
by a factor of a few) is striking.

\floattable
\begin{deluxetable}{lcccrcccrcc}
\tablewidth{0pt}
\tablecaption{Disk Parameter Fits\label{tab:fits}}
\tablehead{\colhead{Object} & \colhead{$\alpha_{\rm CCF}$} & \colhead{$\beta_{\rm CCF}$} & \colhead{$\chi^2_{\rm CCF}/{\rm dof}$} & \colhead{$\rho_{\rm CCF}$} & \colhead{$\alpha_{\tt JAV}$} & \colhead{$\beta_{\tt JAV}$} & \colhead{$\chi^2_{\tt JAV}/{\rm dof}$} & \colhead{$\rho_{\tt JAV}$} & \colhead{Thin Disk} & \colhead{NGC\,5548 Rescaled}\\
&\colhead{(light days)}& & & & \colhead{(light days)}& & & &\colhead{(light days)}&\colhead{(light days)}\\
\colhead{(1)}&\colhead{(2)}&\colhead{(3)}&\colhead{(4)}&\colhead{(5)}&\colhead{(6)}&\colhead{(7)}&\colhead{(8)}&\colhead{(9)}&\colhead{(10)}&\colhead{(11)}
}
\startdata
MCG+08-11-011 & $1.13 \pm 0.31$ & 4/3 & 3.16 & \dots & $1.00 \pm 0.11$ & 4/3 & 16.8 & \dots & 0.35 & 1.52 \\
no 5100\,\AA & $1.24 \pm 0.32$ & 4/3 & 1.72 & \dots & $1.15 \pm 0.11$ & 4/3 & 0.75 & \dots &  &  \\
no {\it u} & $1.15 \pm 0.32$ & 4/3 & 3.91 & \dots & $0.99 \pm 0.11$ & 4/3 & 22.26 & \dots &  &  \\
no {\it u}/5100\,\AA & $1.26 \pm 0.32$ & 4/3 & 2.23 & \dots & $1.15 \pm 0.11$ & 4/3 & 1.01 & \dots &  &  \\
\hline
NGC 2617 & $0.65 \pm 0.32$ & 4/3 & 0.27 & \dots & $0.38 \pm 0.11$ & 4/3 & 2.31 & \dots & 0.22 & 0.96 \\
no {\it Swift} & $0.71 \pm 0.47$ & 4/3 & 0.27 & \dots & $0.51 \pm 0.13$ & 4/3 & 2.66 & \dots &  &  \\
no {\it u}/{\it Swift} & $0.79 \pm 0.48$ & 4/3 & 0.17 & \dots & $0.61 \pm 0.14$ & 4/3 & 0.79 & \dots &  &  \\
no {\it ugriz} & $0.60 \pm 0.44$ & 4/3 & 0.26 & \dots & $0.05 \pm 0.21$ & 4/3 & 1.67 & \dots &  &  \\
no {\it ugriz}/{\it Swift u} & $0.75 \pm 0.46$ & 4/3 & 0.08 & \dots & $0.19 \pm 0.22$ & 4/3 & 0.97 & \dots &  &  \\
no {\it u}/{\it Swift u} & $0.77 \pm 0.34$ & 4/3 & 0.12 & \dots & $0.50 \pm 0.12$ & 4/3 & 1.16 & \dots &  &  \\
\hline
MCG+08-11-011 & $1.86 \pm 6.34$ & $0.92 \pm 2.47$ & 3.94 & $-$1.0 & $0.20 \pm 0.18$ & $3.38 \pm 1.33$ & 21.07 & $-$1.0 &  &  \\
no 5100\,\AA & $\infty$ & $0.00 \pm 99.9$ & 2.05 & 0.0 &  $\infty$ & $0.14 \pm 0.90$ & 0.32 & $-$1.0 &  &  \\
no {\it u} & $\infty$ & $0.00 \pm 3.81$ & 5.10 & $-$1.0 & $0.16 \pm 0.16$ & $3.70 \pm 1.44$ & 30.69 & $-$1.0 &  &  \\
no {\it u}/5100\,\AA & $\infty$ & $0.00 \pm 99.9$ & 2.56 & $-$1.0 & $\infty$ & $0.00 \pm 99.9$ & 0.40 & 0.0 &  &  \\
\hline
NGC 2617 & $0.62 \pm 0.64$ & $1.55 \pm 1.44$ & 0.29 & $-$0.9 & $0.09 \pm 0.12$ & $3.84 \pm 2.07$ & 2.01 & $-$1.0 &  &  \\
no {\it Swift} & $0.23 \pm 1.25$ & $2.90 \pm 8.02$ & 0.31 & $-$1.0 & $0.15 \pm 0.26$ & $3.12 \pm 2.60$ & 2.99 & $-$1.0 &  &  \\
no {\it u}/{\it Swift} & $\infty$ & $0.00 \pm 99.9$ & 0.19 & 0.0 & $\infty$ & $0.00 \pm 99.9$ & 0.61 & 0.0 &  &  \\
no {\it ugriz} & $\infty$ & $0.00 \pm 99.9$ & 0.25 & 0.0 & $\infty$ & $0.00 \pm 7.16$ & 2.02 & $-$1.0 &  &  \\
no {\it ugriz}/{\it Swift u} & $\infty$ & $0.00 \pm 99.9$ & 0.09 & 0.0 & $\infty$ & $0.00 \pm 99.9$ & 1.20 & $-$1.0 &  &  \\
no {\it u}/{\it Swift u} & $0.87 \pm 0.84$ & $1.30 \pm 1.23$ & 0.14 & $-$0.9 & $0.28 \pm 0.21$ & $2.28 \pm 1.07$ & 1.13 & $-$1.0 &  &  \\
\enddata
\tablecomments{Column 1 shows the excluded lags for each fit (where
  the AGN designations are given, all lags were used).  Columns 2 and
  3 give the parameters fit to the ICCF lags, Column 4 gives the
  reduced $\chi^2$, and Column 5 gives the correlation coefficient
  between $\alpha$ and $\beta$.  Columns 6--9 are the same as Columns
  2--5 but for the {\tt JAVELIN} lags.  In some cases, the data prefer
  a flat relation $\beta = 0$, which does not provide a constraint on
  the disk size $\alpha$ (marked as $\infty$ here).  Column 10 gives
  the prediction from thin disk theory
  (Equation\,\ref{equ:tau_0_scale}), and Column 11 gives the fit to
  NGC\,5548 from \citet{Fausnaugh2016} rescaled to the mass and
  mass-accretion rate of these objects (see \S5.3).  }
\end{deluxetable}

\section{Summary}
We have detected inter-band continuum lags in two Seyfert 1 galaxies,
MCG+08-11-011 and NCG\,2617.  This adds one new object to the previous
sample of two AGN with secure measurements of accretion disk
reverberation signals.  We also compared the lags for NGC\,2617 in
2014 to lags measured one year prior in 2013 by \citet{Shappee2014}.
\begin{enumerate}[i.]
\item We generally find longer lags at longer wavelengths, consistent
  with disk reprocessing models.  The exceptions are the \usdss\ data
  for in NGC\,2617 and the \g\ for MCG+08-11-011.  For NGC\,2617,
  these longer lags are probably due to contamination by the Balmer
  continuum.  The origin of the anomalous lag in MCG+08-11-011 is less
  clear, but it may be caused by a similar bias in the broad-band
  filters that is not present in the spectroscopic continuum light
  curve.
\item The X-ray to UV/optical lag in NGC\,2617 is $\sim$2.6 days, and
  there is substantially more structure in the X-ray light curve than
  the UV/optical light curves.  This is inconsistent with standard
  reprocessing models where the X-ray emitting corona directly
  irradiates the surrounding accretion disk.  However, there is still
  some correlation between these light curves, suggesting a
  complicated relationship between the X-ray and UV/optical emission.

\item The lag-wavelength relations (for the UV/optical light curves)
  are consistent with the predictions for reprocessing in a standard
  geometrically thin disk.  However, the inferred disk sizes are
  larger than these predictions by a factor of 3.3 in MCG+08-11-011
  and 2.3 in NGC\,2617.

\item These results may indicate that the observed optical
  luminosities underestimate the total energy generation (mass
  accretion) rates.  Using the {\tt CREAM} physical reprocessing model
  to fit the light curves, we find Eddington ratios larger than would
  be estimated from the optical luminosity by a factor of 4.3 in
  MCG+08-11-011 and a factor of 1.6 in NGC\,2617.  However, these
  differences are not statistically significant, considering
  uncertainty in the SMBH masses.
\end{enumerate}

These results add to the growing body of evidence for additional
structure and possibly physical processes in AGN accretion disks: the X-ray
phenomenology indicates a more complicated situation than simple disk
reprocessing, and there is tension between the sizes of the disks and
standard models.  However, the significance of this statement is
limited by uncertainties in the SMBH masses.  A direct means of
expanding our work is to improve the SMBH mass uncertainties, for
example, by using the dynamical models of \citet{Pancoast2014b} to
estimate the individual virial factor $f$ of each object.  We will
pursue such an analysis for MCG+08-11-011 and NGC\,2617 in future work
using our spectroscopic RM data (see \citealt{Fausnaugh2017b} for
details).  An alternative tactic, which has been adopted by the LCO
AGN Key project, is to expand the sample size of RM-measured AGN disk
sizes.  This will improve the uncertainty on the mean AGN disk size,
and potentially pin down departures from standard models.  Finally,
microlensing of strongly lensed quasars is the only other practical
means of probing the accretion disks around SMBHs.  Both microlensing
and RM find similar disk sizes (e.g., \citealt{Mosquera2013}), and a
thorough and systematic comparison of the microlensing and RM results
is therefore warranted.

M.M.F. acknowledges financial support from a Presidential Fellowship
awarded by The Ohio State University Graduate School. NSF grant
AST-1008882 supported M.M.F., G.D.R., B.M.P., and R.W.P.,
M.C.B. gratefully acknowledges support through NSF CAREER grant
AST-1253702 to Georgia State University.  K.D.D. is supported by an
NSF AAPF fellowship awarded under NSF grant AST-1302093.  C.S.K. is
supported by NSF grant AST-1515876.  K.H. acknowledges support from
STFC grant ST/M001296/1.  This material is based in part upon work
supported by the National Science Foundation (NSF) Graduate Research
Fellowship Program under Grant No.\ DGE-0822215, awarded to C.B.H.
A.M.M. acknowledges the support of NSF grant AST-1211146.  M.E. thanks
the members of the Center for Relativistic Astrophysics at Georgia
Tech, where he was based during the observing campaign, for their warm
hospitality.  J.S.S. acknowledges CNPq, National Council for
Scientific and Technological Development, Brazil.  J.T. acknowledges
support from NSF grant AST-1411685.  Work by S.V.Jr. is supported by
the National Science Foundation Graduate Research Fellowship under
Grant No. DGE-1343012.  Work by W.Z. was supported by NSF grant
AST-1516842.  T.W.-S.H. is supported by the DOE Computational Science
Graduate Fellowship, grant number DE-FG02-97ER25308.  E.R.C. and
S.M.C. gratefully acknowledge the receipt of research grants from the
National Research Foundation (NRF) of South Africa. T.T. acknowledges
support by the National Science Foundation through grant AST-1412315
"Collaborative Research: New Frontiers in Reverberation Mapping," and
by the Packard Foundation through a Packard Research Fellowship.
D.J.S.  acknowledges support from NSF grants AST-1412504 and
AST-1517649.  A.J.B. and L.P. have been supported by NSF grant
AST-1412693.  B.J.S. is supported by NASA through Hubble Fellowship
grant HF-51348.001 awarded by the Space Telescope Science Institute
that is operated by the Association of Universities for Research in
Astronomy, Inc., for NASA, under contract NAS 5-26555.

MCB acknowledges support through grant {\it HST} GO-13816 from the
Space Telescope Science Institute, which is operated by the
Association of Universities for Research in Astronomy, Inc., under
NASA contract NAS5-26555.  This research has made use of the XRT Data
Analysis Software (XRTDAS) developed under the responsibility of the
ASI Science Data Center (ASDC), Italy.  This work is based on
observations obtained at the MDM Observatory, operated by Dartmouth
College, Columbia University, Ohio State University, Ohio University,
and the University of Michigan.  This paper is partly based on
observations collected at the Wise Observatory with the C18
telescope. The C18 telescope and most of its equipment were acquired
with a grant from the Israel Space Agency (ISA) to operate a
Near-Earth Asteroid Knowledge Center at Tel Aviv University.  The
Fountainwood Observatory would like to thank the HHMI for its support
of science research for undergraduate students at Southwestern
University.  This research has made use of NASA's Astrophysics Data
System, as well as the NASA/IPAC Extragalactic Database (NED) which is
operated by the Jet Propulsion Laboratory, California Institute of
Technology, under contract with the National Aeronautics and Space
Administration.

\facility{McGraw-Hill, HST, Wise Observatory, Fountainwood Observatory
  BYU:0.9m, CrAO:0.7m, WIRO, LCO, LCOGT, SSO:1m, Swift}

\software{ Astropy \citep{astropy},
  IRAF \citep{iraf},
  Matplotlib \citep{matplotlib},
  Numpy \citep{numpy},
  Scipy \citep{scipy} }

\bibliography{ms.bbl}

\begin{thebibliography}{}
\expandafter\ifx\csname natexlab\endcsname\relax\def\natexlab#1{#1}\fi

\bibitem[{{Abazajian} {et~al.}(2009){Abazajian}, {Adelman-McCarthy},
  {Ag{\"u}eros}, {Allam}, {Allende Prieto}, {An}, {Anderson}, {Anderson},
  {Annis}, {Bahcall}, \& et~al.}]{Abazajian2009}
{Abazajian}, K.~N., {Adelman-McCarthy}, J.~K., {Ag{\"u}eros}, M.~A., {et~al.}
  2009, \apjs, 182, 543

\bibitem[{{Abramowicz} {et~al.}(1988){Abramowicz}, {Czerny}, {Lasota}, \&
  {Szuszkiewicz}}]{Abramowicz1988}
{Abramowicz}, M.~A., {Czerny}, B., {Lasota}, J.~P., \& {Szuszkiewicz}, E. 1988,
  \apj, 332, 646

\bibitem[{{Alard} \& {Lupton}(1998)}]{Alard1998}
{Alard}, C., \& {Lupton}, R.~H. 1998, \apj, 503, 325

\bibitem[{{Ar{\'e}valo} {et~al.}(2008){Ar{\'e}valo}, {Uttley}, {Kaspi},
  {Breedt}, {Lira}, \& {McHardy}}]{Arevalo2008}
{Ar{\'e}valo}, P., {Uttley}, P., {Kaspi}, S., {et~al.} 2008, \mnras, 389, 1479

\bibitem[{{Ar{\'e}valo} {et~al.}(2009){Ar{\'e}valo}, {Uttley}, {Lira},
  {Breedt}, {McHardy}, \& {Churazov}}]{Arevalo2009}
{Ar{\'e}valo}, P., {Uttley}, P., {Lira}, P., {et~al.} 2009, \mnras, 397, 2004

\bibitem[{{Astropy Collaboration} {et~al.}(2013){Astropy Collaboration},
  {Robitaille}, {Tollerud}, {Greenfield}, {Droettboom}, {Bray}, {Aldcroft},
  {Davis}, {Ginsburg}, {Price-Whelan}, {Kerzendorf}, {Conley}, {Crighton},
  {Barbary}, {Muna}, {Ferguson}, {Grollier}, {Parikh}, {Nair}, {Unther},
  {Deil}, {Woillez}, {Conseil}, {Kramer}, {Turner}, {Singer}, {Fox}, {Weaver},
  {Zabalza}, {Edwards}, {Azalee Bostroem}, {Burke}, {Casey}, {Crawford},
  {Dencheva}, {Ely}, {Jenness}, {Labrie}, {Lim}, {Pierfederici}, {Pontzen},
  {Ptak}, {Refsdal}, {Servillat}, \& {Streicher}}]{astropy}
{Astropy Collaboration}, {Robitaille}, T.~P., {Tollerud}, E.~J., {et~al.} 2013,
  \aap, 558, A33

\bibitem[{{Balbus} \& {Hawley}(1998)}]{Balbus1998}
{Balbus}, S.~A., \& {Hawley}, J.~F. 1998, Reviews of Modern Physics, 70, 1

\bibitem[{{Begelman} {et~al.}(2015){Begelman}, {Armitage}, \&
  {Reynolds}}]{Begelman2015}
{Begelman}, M.~C., {Armitage}, P.~J., \& {Reynolds}, C.~S. 2015, \apj, 809, 118

\bibitem[{{Begelman} \& {Silk}(2017)}]{Begelman2017}
{Begelman}, M.~C., \& {Silk}, J. 2017, \mnras, 464, 2311

\bibitem[{{Blackburne} {et~al.}(2011){Blackburne}, {Pooley}, {Rappaport}, \&
  {Schechter}}]{Blackburne2011}
{Blackburne}, J.~A., {Pooley}, D., {Rappaport}, S., \& {Schechter}, P.~L. 2011,
  \apj, 729, 34

\bibitem[{{Blaes}(2014)}]{Blaes2014}
{Blaes}, O. 2014, \ssr, 183, 21

\bibitem[{{Blandford} \& {McKee}(1982)}]{Blandford1982}
{Blandford}, R.~D., \& {McKee}, C.~F. 1982, \apj, 255, 419

\bibitem[{{Breedt} {et~al.}(2009){Breedt}, {Ar{\'e}valo}, {McHardy}, {Uttley},
  {Sergeev}, {Minezaki}, {Yoshii}, {Gaskell}, {Cackett}, {Horne}, \&
  {Koshida}}]{Breedt2009}
{Breedt}, E., {Ar{\'e}valo}, P., {McHardy}, I.~M., {et~al.} 2009, \mnras, 394,
  427

\bibitem[{{Breedt} {et~al.}(2010){Breedt}, {McHardy}, {Ar{\'e}valo}, {Uttley},
  {Sergeev}, {Minezaki}, {Yoshii}, {Sakata}, {Lira}, \& {Chesnok}}]{Breedt2010}
{Breedt}, E., {McHardy}, I.~M., {Ar{\'e}valo}, P., {et~al.} 2010, \mnras, 403,
  605

\bibitem[{{Brosch} {et~al.}(2008){Brosch}, {Polishook}, {Shporer}, {Kaspi},
  {Berwald}, \& {Manulis}}]{Brosch2008}
{Brosch}, N., {Polishook}, D., {Shporer}, A., {et~al.} 2008, \apss, 314, 163

\bibitem[{{Brown} {et~al.}(2013){Brown}, {Baliber}, {Bianco}, {Bowman},
  {Burleson}, {Conway}, {Crellin}, {Depagne}, {De Vera}, {Dilday}, {Dragomir},
  {Dubberley}, {Eastman}, {Elphick}, {Falarski}, {Foale}, {Ford}, {Fulton},
  {Garza}, {Gomez}, {Graham}, {Greene}, {Haldeman}, {Hawkins}, {Haworth},
  {Haynes}, {Hidas}, {Hjelstrom}, {Howell}, {Hygelund}, {Lister}, {Lobdill},
  {Martinez}, {Mullins}, {Norbury}, {Parrent}, {Paulson}, {Petry}, {Pickles},
  {Posner}, {Rosing}, {Ross}, {Sand}, {Saunders}, {Shobbrook}, {Shporer},
  {Street}, {Thomas}, {Tsapras}, {Tufts}, {Valenti}, {Vander Horst}, {Walker},
  {White}, \& {Willis}}]{Brown2013}
{Brown}, T.~M., {Baliber}, N., {Bianco}, F.~B., {et~al.} 2013, \pasp, 125, 1031

\bibitem[{{Buisson} {et~al.}(2017){Buisson}, {Lohfink}, {Alston}, \&
  {Fabian}}]{Buisson2017}
{Buisson}, D.~J.~K., {Lohfink}, A.~M., {Alston}, W.~N., \& {Fabian}, A.~C.
  2017, \mnras, 464, 3194

\bibitem[{{Burbidge}(1967)}]{Burbidge1967}
{Burbidge}, E.~M. 1967, \araa, 5, 399

\bibitem[{{Burrows} {et~al.}(2005){Burrows}, {Hill}, {Nousek}, {Kennea},
  {Wells}, {Osborne}, {Abbey}, {Beardmore}, {Mukerjee}, {Short}, {Chincarini},
  {Campana}, {Citterio}, {Moretti}, {Pagani}, {Tagliaferri}, {Giommi},
  {Capalbi}, {Tamburelli}, {Angelini}, {Cusumano}, {Br{\"a}uninger}, {Burkert},
  \& {Hartner}}]{Burrows2005}
{Burrows}, D.~N., {Hill}, J.~E., {Nousek}, J.~A., {et~al.} 2005, \ssr, 120, 165

\bibitem[{{Cackett} {et~al.}(2007){Cackett}, {Horne}, \&
  {Winkler}}]{Cackett2007}
{Cackett}, E.~M., {Horne}, K., \& {Winkler}, H. 2007, \mnras, 380, 669

\bibitem[{{Capellupo} {et~al.}(2015){Capellupo}, {Netzer}, {Lira},
  {Trakhtenbrot}, \& {Mej{\'{\i}}a-Restrepo}}]{Capellupo2015}
{Capellupo}, D.~M., {Netzer}, H., {Lira}, P., {Trakhtenbrot}, B., \&
  {Mej{\'{\i}}a-Restrepo}, J. 2015, \mnras, 446, 3427

\bibitem[{{Collier} {et~al.}(1998){Collier}, {Horne}, {Kaspi}, {Netzer},
  {Peterson}, {Wanders}, {Alexander}, {Bertram}, {Comastri}, {Gaskell},
  {Malkov}, {Maoz}, {Mignoli}, {Pogge}, {Pronik}, {Sergeev}, {Snedden},
  {Stirpe}, {Bochkarev}, {Burenkov}, {Shapovalova}, \& {Wagner}}]{Collier1998}
{Collier}, S.~J., {Horne}, K., {Kaspi}, S., {et~al.} 1998, \apj, 500, 162

\bibitem[{{Davidson} \& {Netzer}(1979)}]{Davidson1979}
{Davidson}, K., \& {Netzer}, H. 1979, Reviews of Modern Physics, 51, 715

\bibitem[{{De Rosa} {et~al.}(2015){De Rosa}, {Peterson}, {Ely}, {Kriss},
  {Crenshaw}, {Horne}, {Korista}, {Netzer}, {Pogge}, {Ar{\'e}valo}, {Barth},
  {Bentz}, {Brandt}, {Breeveld}, {Brewer}, {Dalla Bont{\`a}}, {De
  Lorenzo-C{\'a}ceres}, {Denney}, {Dietrich}, {Edelson}, {Evans}, {Fausnaugh},
  {Gehrels}, {Gelbord}, {Goad}, {Grier}, {Grupe}, {Hall}, {Kaastra}, {Kelly},
  {Kennea}, {Kochanek}, {Lira}, {Mathur}, {McHardy}, {Nousek}, {Pancoast},
  {Papadakis}, {Pei}, {Schimoia}, {Siegel}, {Starkey}, {Treu}, {Uttley},
  {Vaughan}, {Vestergaard}, {Villforth}, {Yan}, {Young}, \& {Zu}}]{DeRosa2015}
{De Rosa}, G., {Peterson}, B.~M., {Ely}, J., {et~al.} 2015, \apj, 806, 128

\bibitem[{{Edelson} {et~al.}(2015){Edelson}, {Gelbord}, {Horne}, {McHardy},
  {Peterson}, {Ar{\'e}valo}, {Breeveld}, {De Rosa}, {Evans}, {Goad}, {Kriss},
  {Brandt}, {Gehrels}, {Grupe}, {Kennea}, {Kochanek}, {Nousek}, {Papadakis},
  {Siegel}, {Starkey}, {Uttley}, {Vaughan}, {Young}, {Barth}, {Bentz},
  {Brewer}, {Crenshaw}, {Dalla Bont{\`a}}, {De Lorenzo-C{\'a}ceres}, {Denney},
  {Dietrich}, {Ely}, {Fausnaugh}, {Grier}, {Hall}, {Kaastra}, {Kelly},
  {Korista}, {Lira}, {Mathur}, {Netzer}, {Pancoast}, {Pei}, {Pogge},
  {Schimoia}, {Treu}, {Vestergaard}, {Villforth}, {Yan}, \& {Zu}}]{Edelson2015}
{Edelson}, R., {Gelbord}, J.~M., {Horne}, K., {et~al.} 2015, \apj, 806, 129

\bibitem[{{Edelson} {et~al.}(2017){Edelson}, {Gelbord}, {Cackett}, {Connolly},
  {Done}, {Fausnaugh}, {Gardner}, {Gehrels}, {Goad}, {Horne}, {McHardy},
  {Peterson}, {Vaughan}, {Vestergaard}, {Breeveld}, {Barth}, {Bentz},
  {Bottorff}, {Brandt}, {Crawford}, {Dalla Bont{\`a}}, {Emmanoulopoulos},
  {Evans}, {Figuera Jaimes}, {Filippenko}, {Ferland}, {Grupe}, {Joner},
  {Kennea}, {Korista}, {Krimm}, {Kriss}, {Leonard}, {Mathur}, {Netzer},
  {Nousek}, {Page}, {Romero-Colmenero}, {Siegel}, {Starkey}, {Treu}, {Vogler},
  {Winkler}, \& {Zheng}}]{Edelson2017}
{Edelson}, R., {Gelbord}, J., {Cackett}, E., {et~al.} 2017, \apj, 840, 41

\bibitem[{{Elvis} {et~al.}(1994){Elvis}, {Wilkes}, {McDowell}, {Green},
  {Bechtold}, {Willner}, {Oey}, {Polomski}, \& {Cutri}}]{Elvis1994}
{Elvis}, M., {Wilkes}, B.~J., {McDowell}, J.~C., {et~al.} 1994, \apjs, 95, 1

\bibitem[{{Fausnaugh}(2017)}]{Fausnaugh2017}
{Fausnaugh}, M.~M. 2017, \pasp, 129, 024007

\bibitem[{{Fausnaugh} {et~al.}(2016){Fausnaugh}, {Denney}, {Barth}, {Bentz},
  {Bottorff}, {Carini}, {Croxall}, {De Rosa}, {Goad}, {Horne}, {Joner},
  {Kaspi}, {Kim}, {Klimanov}, {Kochanek}, {Leonard}, {Netzer}, {Peterson},
  {Schn{\"u}lle}, {Sergeev}, {Vestergaard}, {Zheng}, {Zu}, {Anderson},
  {Ar{\'e}valo}, {Bazhaw}, {Borman}, {Boroson}, {Brandt}, {Breeveld}, {Brewer},
  {Cackett}, {Crenshaw}, {Dalla Bont{\`a}}, {De Lorenzo-C{\'a}ceres},
  {Dietrich}, {Edelson}, {Efimova}, {Ely}, {Evans}, {Filippenko}, {Flatland},
  {Gehrels}, {Geier}, {Gelbord}, {Gonzalez}, {Gorjian}, {Grier}, {Grupe},
  {Hall}, {Hicks}, {Horenstein}, {Hutchison}, {Im}, {Jensen}, {Jones},
  {Kaastra}, {Kelly}, {Kennea}, {Kim}, {Korista}, {Kriss}, {Lee}, {Lira},
  {MacInnis}, {Manne-Nicholas}, {Mathur}, {McHardy}, {Montouri}, {Musso},
  {Nazarov}, {Norris}, {Nousek}, {Okhmat}, {Pancoast}, {Papadakis}, {Parks},
  {Pei}, {Pogge}, {Pott}, {Rafter}, {Rix}, {Saylor}, {Schimoia}, {Siegel},
  {Spencer}, {Starkey}, {Sung}, {Teems}, {Treu}, {Turner}, {Uttley},
  {Villforth}, {Weiss}, {Woo}, {Yan}, \& {Young}}]{Fausnaugh2016}
{Fausnaugh}, M.~M., {Denney}, K.~D., {Barth}, A.~J., {et~al.} 2016, \apj, 821,
  56

\bibitem[{{Fausnaugh} {et~al.}(2017){Fausnaugh}, {Grier}, {Bentz}, {Denney},
  {De Rosa}, {Peterson}, {Kochanek}, {Pogge}, {Adams}, {Barth}, {Beatty},
  {Bhattacharjee}, {Borman}, {Boroson}, {Bottorff}, {Brown}, {Brown},
  {Brotherton}, {Coker}, {Crawford}, {Croxall}, {Eftekharzadeh}, {Eracleous},
  {Joner}, {Henderson}, {Holoien}, {Horne}, {Hutchison}, {Kaspi}, {Kim},
  {King}, {Li}, {Lochhaas}, {Ma}, {MacInnis}, {Manne-Nicholas}, {Mason},
  {Montuori}, {Mosquera}, {Mudd}, {Musso}, {Nazarov}, {Nguyen}, {Okhmat},
  {Onken}, {Ou-Yang}, {Pancoast}, {Pei}, {Penny}, {Poleski}, {Rafter},
  {Romero-Colmenero}, {Runnoe}, {Sand}, {Schimoia}, {Sergeev}, {Shappee},
  {Simonian}, {Somers}, {Spencer}, {Starkey}, {Stevens}, {Tayar}, {Treu},
  {Valenti}, {Van Saders}, {Villanueva}, {Villforth}, {Weiss}, {Winkler}, \&
  {Zhu}}]{Fausnaugh2017b}
{Fausnaugh}, M.~M., {Grier}, C.~J., {Bentz}, M.~C., {et~al.} 2017, \apj, 840,
  97

\bibitem[{{Gardner} \& {Done}(2017)}]{Gardner2017}
{Gardner}, E., \& {Done}, C. 2017, \mnras, 470, 3591

\bibitem[{{Gaskell} \& {Peterson}(1987)}]{Gaskell1987}
{Gaskell}, C.~M., \& {Peterson}, B.~M. 1987, \apjs, 65, 1

\bibitem[{{Giustini} {et~al.}(2017){Giustini}, {Costantini}, {De Marco},
  {Svoboda}, {Motta}, {Proga}, {Saxton}, {Ferrigno}, {Longinotti}, {Miniutti},
  {Grupe}, {Mathur}, {Shappee}, {Prieto}, \& {Stanek}}]{Giustini2017}
{Giustini}, M., {Costantini}, E., {De Marco}, B., {et~al.} 2017, \aap, 597, A66

\bibitem[{{Gliozzi} {et~al.}(2017){Gliozzi}, {Papadakis}, {Grupe}, {Brinkmann},
  \& {R{\"a}th}}]{Gliozzi2017}
{Gliozzi}, M., {Papadakis}, I.~E., {Grupe}, D., {Brinkmann}, W.~P., \&
  {R{\"a}th}, C. 2017, \mnras, 464, 3955

\bibitem[{{Goad} \& {Korista}(2014)}]{Goad2014}
{Goad}, M.~R., \& {Korista}, K.~T. 2014, \mnras, 444, 43

\bibitem[{{Haardt} \& {Maraschi}(1991)}]{Haardt1991}
{Haardt}, F., \& {Maraschi}, L. 1991, \apjl, 380, L51

\bibitem[{{Horne}(1994)}]{Horne1994}
{Horne}, K. 1994, in Astronomical Society of the Pacific Conference Series,
  Vol.~69, Reverberation Mapping of the Broad-Line Region in Active Galactic
  Nuclei, ed. P.~M. {Gondhalekar}, K.~{Horne}, \& B.~M. {Peterson}, 23

\bibitem[{Hunter(2007)}]{matplotlib}
Hunter, J.~D. 2007, Computing in Science {\&} Engineering, 9, 90

\bibitem[{{Jiang} {et~al.}(2014){Jiang}, {Stone}, \& {Davis}}]{Jiang2014b}
{Jiang}, Y.-F., {Stone}, J.~M., \& {Davis}, S.~W. 2014, \apj, 784, 169

\bibitem[{{Jiang} {et~al.}(2017){Jiang}, {Green}, {Greene}, {Morganson},
  {Shen}, {Pancoast}, {MacLeod}, {Anderson}, {Brandt}, {Grier}, {Rix}, {Ruan},
  {Protopapas}, {Scott}, {Burgett}, {Hodapp}, {Huber}, {Kaiser}, {Kudritzki},
  {Magnier}, {Metcalfe}, {Tonry}, {Wainscoat}, \& {Waters}}]{Jiang2017}
{Jiang}, Y.-F., {Green}, P.~J., {Greene}, J.~E., {et~al.} 2017, \apj, 836, 186

\bibitem[{{Jim{\'e}nez-Vicente} {et~al.}(2014){Jim{\'e}nez-Vicente},
  {Mediavilla}, {Kochanek}, {Mu{\~n}oz}, {Motta}, {Falco}, \&
  {Mosquera}}]{Jimenez-Vicente2014}
{Jim{\'e}nez-Vicente}, J., {Mediavilla}, E., {Kochanek}, C.~S., {et~al.} 2014,
  \apj, 783, 47

\bibitem[{{Kishimoto} {et~al.}(2008){Kishimoto}, {Antonucci}, {Blaes},
  {Lawrence}, {Boisson}, {Albrecht}, \& {Leipski}}]{Kishimoto2008}
{Kishimoto}, M., {Antonucci}, R., {Blaes}, O., {et~al.} 2008, \nat, 454, 492

\bibitem[{{Kishimoto} {et~al.}(2004){Kishimoto}, {Antonucci}, {Boisson}, \&
  {Blaes}}]{Kishimoto2004}
{Kishimoto}, M., {Antonucci}, R., {Boisson}, C., \& {Blaes}, O. 2004, \mnras,
  354, 1065

\bibitem[{{Kokubo}(2015)}]{Kokubo2015}
{Kokubo}, M. 2015, \mnras, 449, 94

\bibitem[{{Kokubo}(2016)}]{Kokubo2016}
---. 2016, \pasj, 68, 52

\bibitem[{{Korista} \& {Goad}(2001)}]{Korista2001}
{Korista}, K.~T., \& {Goad}, M.~R. 2001, \apj, 553, 695

\bibitem[{{Koz{\l}owski}(2017)}]{Kozlowski2017}
{Koz{\l}owski}, S. 2017, \aap, 597, A128

\bibitem[{{Krolik} {et~al.}(1991){Krolik}, {Horne}, {Kallman}, {Malkan},
  {Edelson}, \& {Kriss}}]{Krolik1991}
{Krolik}, J.~H., {Horne}, K., {Kallman}, T.~R., {et~al.} 1991, \apj, 371, 541

\bibitem[{{LaMassa} {et~al.}(2015){LaMassa}, {Cales}, {Moran}, {Myers},
  {Richards}, {Eracleous}, {Heckman}, {Gallo}, \& {Urry}}]{LaMassa2015}
{LaMassa}, S.~M., {Cales}, S., {Moran}, E.~C., {et~al.} 2015, \apj, 800, 144

\bibitem[{{Lira} {et~al.}(2015){Lira}, {Ar{\'e}valo}, {Uttley}, {McHardy}, \&
  {Videla}}]{Lira2015}
{Lira}, P., {Ar{\'e}valo}, P., {Uttley}, P., {McHardy}, I.~M.~M., \& {Videla},
  L. 2015, \mnras, 454, 368

\bibitem[{{Luo} \& {Liang}(1998)}]{Luo1998}
{Luo}, C., \& {Liang}, E.~P. 1998, \apj, 498, 307

\bibitem[{{MacLeod} {et~al.}(2016){MacLeod}, {Ross}, {Lawrence}, {Goad},
  {Horne}, {Burgett}, {Chambers}, {Flewelling}, {Hodapp}, {Kaiser}, {Magnier},
  {Wainscoat}, \& {Waters}}]{MacLeod2016}
{MacLeod}, C.~L., {Ross}, N.~P., {Lawrence}, A., {et~al.} 2016, \mnras, 457,
  389

\bibitem[{{Maoz} {et~al.}(2002){Maoz}, {Markowitz}, {Edelson}, \&
  {Nandra}}]{Maoz2002}
{Maoz}, D., {Markowitz}, A., {Edelson}, R., \& {Nandra}, K. 2002, \aj, 124,
  1988

\bibitem[{{Marshall} {et~al.}(2008){Marshall}, {Ryle}, \&
  {Miller}}]{Marshall2008}
{Marshall}, K., {Ryle}, W.~T., \& {Miller}, H.~R. 2008, \apj, 677, 880

\bibitem[{{McHardy} {et~al.}(2014){McHardy}, {Cameron}, {Dwelly}, {Connolly},
  {Lira}, {Emmanoulopoulos}, {Gelbord}, {Breedt}, {Arevalo}, \&
  {Uttley}}]{Mchardy2014}
{McHardy}, I.~M., {Cameron}, D.~T., {Dwelly}, T., {et~al.} 2014, \mnras, 444,
  1469

\bibitem[{{McHardy} {et~al.}(2016){McHardy}, {Connolly}, {Peterson}, {Bieryla},
  {Chand}, {Elvis}, {Emmanoulopoulos}, {Falco}, {Gandhi}, {Kaspi}, {Latham},
  {Lira}, {McCully}, {Netzer}, \& {Uemura}}]{McHardy2016}
{McHardy}, I.~M., {Connolly}, S.~D., {Peterson}, B.~M., {et~al.} 2016,
  Astronomische Nachrichten, 337, 500

\bibitem[{{Morgan} {et~al.}(2010){Morgan}, {Kochanek}, {Morgan}, \&
  {Falco}}]{Morgan2010}
{Morgan}, C.~W., {Kochanek}, C.~S., {Morgan}, N.~D., \& {Falco}, E.~E. 2010,
  \apj, 712, 1129

\bibitem[{{Mosquera} {et~al.}(2013){Mosquera}, {Kochanek}, {Chen}, {Dai},
  {Blackburne}, \& {Chartas}}]{Mosquera2013}
{Mosquera}, A.~M., {Kochanek}, C.~S., {Chen}, B., {et~al.} 2013, \apj, 769, 53

\bibitem[{{Narayan} \& {Yi}(1995)}]{Narayan1995}
{Narayan}, R., \& {Yi}, I. 1995, \apj, 452, 710

\bibitem[{{Nealon} {et~al.}(2015){Nealon}, {Price}, \& {Nixon}}]{Nealon2015}
{Nealon}, R., {Price}, D.~J., \& {Nixon}, C.~J. 2015, \mnras, 448, 1526

\bibitem[{{Nenkova} {et~al.}(2008){Nenkova}, {Sirocky}, {Nikutta},
  {Ivezi{\'c}}, \& {Elitzur}}]{Nenkova2008}
{Nenkova}, M., {Sirocky}, M.~M., {Nikutta}, R., {Ivezi{\'c}}, {\v Z}., \&
  {Elitzur}, M. 2008, \apj, 685, 160

\bibitem[{{Noble} {et~al.}(2011){Noble}, {Krolik}, {Schnittman}, \&
  {Hawley}}]{Noble2011}
{Noble}, S.~C., {Krolik}, J.~H., {Schnittman}, J.~D., \& {Hawley}, J.~F. 2011,
  \apj, 743, 115

\bibitem[{{Oknyansky} {et~al.}(2017){Oknyansky}, {Gaskell}, {Huseynov},
  {Lipunov}, {Shatsky}, {Tsygankov}, {Gorbovskoy}, {Mikailov}, {Tatarnikov},
  {Buckley}, {Metlov}, {Nadzhip}, {Kuznetsov}, {Balanutza}, {Burlak},
  {Galazutdinov}, {Artamonov}, {Salmanov}, {Malanchev}, \&
  {Oknyansky}}]{Oknyansky2017}
{Oknyansky}, V.~L., {Gaskell}, C.~M., {Huseynov}, N.~A., {et~al.} 2017, \mnras,
  467, 1496

\bibitem[{Oliphant(2007)}]{scipy}
Oliphant, T.~E. 2007, Computing in Science {\&} Engineering, 9, 10

\bibitem[{{Onken} {et~al.}(2004){Onken}, {Ferrarese}, {Merritt}, {Peterson},
  {Pogge}, {Vestergaard}, \& {Wandel}}]{Onken2004}
{Onken}, C.~A., {Ferrarese}, L., {Merritt}, D., {et~al.} 2004, \apj, 615, 645

\bibitem[{{Page} \& {Thorne}(1974)}]{Page1974}
{Page}, D.~N., \& {Thorne}, K.~S. 1974, \apj, 191, 499

\bibitem[{{Pancoast} {et~al.}(2014){Pancoast}, {Brewer}, {Treu}, {Park},
  {Barth}, {Bentz}, \& {Woo}}]{Pancoast2014b}
{Pancoast}, A., {Brewer}, B.~J., {Treu}, T., {et~al.} 2014, \mnras, 445, 3073

\bibitem[{{Pei} {et~al.}(2017){Pei}, {Fausnaugh}, {Barth}, {Peterson}, {Bentz},
  {De Rosa}, {Denney}, {Goad}, {Kochanek}, {Korista}, {Kriss}, {Pogge},
  {Bennert}, {Brotherton}, {Clubb}, {Dalla Bont{\`a}}, {Filippenko}, {Greene},
  {Grier}, {Vestergaard}, {Zheng}, {Adams}, {Beatty}, {Bigley}, {Brown},
  {Brown}, {Canalizo}, {Comerford}, {Coker}, {Corsini}, {Croft}, {Croxall},
  {Deason}, {Eracleous}, {Fox}, {Gates}, {Henderson}, {Holmbeck}, {Holoien},
  {Jensen}, {Johnson}, {Kelly}, {Kim}, {King}, {Lau}, {Li}, {Lochhaas}, {Ma},
  {Manne-Nicholas}, {Mauerhan}, {Malkan}, {McGurk}, {Morelli}, {Mosquera},
  {Mudd}, {Muller Sanchez}, {Nguyen}, {Ochner}, {Ou-Yang}, {Pancoast}, {Penny},
  {Pizzella}, {Poleski}, {Runnoe}, {Scott}, {Schimoia}, {Shappee}, {Shivvers},
  {Simonian}, {Siviero}, {Somers}, {Stevens}, {Strauss}, {Tayar}, {Tejos},
  {Treu}, {Van Saders}, {Vican}, {Villanueva}, {Yuk}, {Zakamska}, {Zhu},
  {Anderson}, {Ar{\'e}valo}, {Bazhaw}, {Bisogni}, {Borman}, {Bottorff},
  {Brandt}, {Breeveld}, {Cackett}, {Carini}, {Crenshaw}, {De
  Lorenzo-C{\'a}ceres}, {Dietrich}, {Edelson}, {Efimova}, {Ely}, {Evans},
  {Ferland}, {Flatland}, {Gehrels}, {Geier}, {Gelbord}, {Grupe}, {Gupta},
  {Hall}, {Hicks}, {Horenstein}, {Horne}, {Hutchison}, {Im}, {Joner}, {Jones},
  {Kaastra}, {Kaspi}, {Kelly}, {Kennea}, {Kim}, {Kim}, {Klimanov}, {Lee},
  {Leonard}, {Lira}, {MacInnis}, {Mathur}, {McHardy}, {Montouri}, {Musso},
  {Nazarov}, {Netzer}, {Norris}, {Nousek}, {Okhmat}, {Papadakis}, {Parks},
  {Pott}, {Rafter}, {Rix}, {Saylor}, {Schn{\"u}lle}, {Sergeev}, {Siegel},
  {Skielboe}, {Spencer}, {Starkey}, {Sung}, {Teems}, {Turner}, {Uttley},
  {Villforth}, {Weiss}, {Woo}, {Yan}, {Young}, \& {Zu}}]{Pei2017}
{Pei}, L., {Fausnaugh}, M.~M., {Barth}, A.~J., {et~al.} 2017, \apj, 837, 131

\bibitem[{{Pereyra} {et~al.}(2006){Pereyra}, {Vanden Berk}, {Turnshek},
  {Hillier}, {Wilhite}, {Kron}, {Schneider}, \& {Brinkmann}}]{Pereyra2006}
{Pereyra}, N.~A., {Vanden Berk}, D.~E., {Turnshek}, D.~A., {et~al.} 2006, \apj,
  642, 87

\bibitem[{{Peterson}(1993)}]{Peterson1993}
{Peterson}, B.~M. 1993, \pasp, 105, 247

\bibitem[{{Peterson}(2014)}]{Peterson2014}
---. 2014, \ssr, 183, 253

\bibitem[{{Peterson} {et~al.}(2004){Peterson}, {Ferrarese}, {Gilbert}, {Kaspi},
  {Malkan}, {Maoz}, {Merritt}, {Netzer}, {Onken}, {Pogge}, {Vestergaard}, \&
  {Wandel}}]{Peterson2004}
{Peterson}, B.~M., {Ferrarese}, L., {Gilbert}, K.~M., {et~al.} 2004, \apj, 613,
  682

\bibitem[{{Rees}(1984)}]{Rees1984}
{Rees}, M.~J. 1984, \araa, 22, 471

\bibitem[{{Reis} \& {Miller}(2013)}]{Reis2013}
{Reis}, R.~C., \& {Miller}, J.~M. 2013, \apjl, 769, L7

\bibitem[{{Reynolds} \& {Nowak}(2003)}]{Reynolds2003}
{Reynolds}, C.~S., \& {Nowak}, M.~A. 2003, \physrep, 377, 389

\bibitem[{{Roming} {et~al.}(2005){Roming}, {Kennedy}, {Mason}, {Nousek}, {Ahr},
  {Bingham}, {Broos}, {Carter}, {Hancock}, {Huckle}, {Hunsberger}, {Kawakami},
  {Killough}, {Koch}, {McLelland}, {Smith}, {Smith}, {Soto}, {Boyd},
  {Breeveld}, {Holland}, {Ivanushkina}, {Pryzby}, {Still}, \&
  {Stock}}]{Roming2005}
{Roming}, P.~W.~A., {Kennedy}, T.~E., {Mason}, K.~O., {et~al.} 2005, \ssr, 120,
  95

\bibitem[{{Runnoe} {et~al.}(2012){Runnoe}, {Brotherton}, \&
  {Shang}}]{Runnoe2012}
{Runnoe}, J.~C., {Brotherton}, M.~S., \& {Shang}, Z. 2012, \mnras, 422, 478

\bibitem[{{Runnoe} {et~al.}(2016){Runnoe}, {Cales}, {Ruan}, {Eracleous},
  {Anderson}, {Shen}, {Green}, {Morganson}, {LaMassa}, {Greene}, {Dwelly},
  {Schneider}, {Merloni}, {Georgakakis}, \& {Roman-Lopes}}]{Runnoe2016}
{Runnoe}, J.~C., {Cales}, S., {Ruan}, J.~J., {et~al.} 2016, \mnras, 455, 1691

\bibitem[{{S{\c a}dowski}(2016)}]{Sadowski2016}
{S{\c a}dowski}, A. 2016, \mnras, 459, 4397

\bibitem[{{S{\c a}dowski} \& {Narayan}(2015)}]{Sadowski2015a}
{S{\c a}dowski}, A., \& {Narayan}, R. 2015, \mnras, 454, 2372

\bibitem[{{S{\c a}dowski} {et~al.}(2014){S{\c a}dowski}, {Narayan}, {McKinney},
  \& {Tchekhovskoy}}]{Sadowski2014}
{S{\c a}dowski}, A., {Narayan}, R., {McKinney}, J.~C., \& {Tchekhovskoy}, A.
  2014, \mnras, 439, 503

\bibitem[{{Schmidt} {et~al.}(2012){Schmidt}, {Rix}, {Shields}, {Knecht},
  {Hogg}, {Maoz}, \& {Bovy}}]{Schmidt2012}
{Schmidt}, K.~B., {Rix}, H.-W., {Shields}, J.~C., {et~al.} 2012, \apj, 744, 147

\bibitem[{{Schnittman} \& {Krolik}(2013)}]{Schnittman2013b}
{Schnittman}, J.~D., \& {Krolik}, J.~H. 2013, \apj, 777, 11

\bibitem[{{Schnittman} {et~al.}(2013){Schnittman}, {Krolik}, \&
  {Noble}}]{Schnittman2013a}
{Schnittman}, J.~D., {Krolik}, J.~H., \& {Noble}, S.~C. 2013, \apj, 769, 156

\bibitem[{{Schnittman} {et~al.}(2016){Schnittman}, {Krolik}, \&
  {Noble}}]{Schnittman2016}
---. 2016, \apj, 819, 48

\bibitem[{{Sergeev} {et~al.}(2005){Sergeev}, {Doroshenko}, {Golubinskiy},
  {Merkulova}, \& {Sergeeva}}]{Sergeev2005}
{Sergeev}, S.~G., {Doroshenko}, V.~T., {Golubinskiy}, Y.~V., {Merkulova},
  N.~I., \& {Sergeeva}, E.~A. 2005, \apj, 622, 129

\bibitem[{{Shakura} \& {Sunyaev}(1973)}]{Shakura1973}
{Shakura}, N.~I., \& {Sunyaev}, R.~A. 1973, \aap, 24, 337

\bibitem[{{Shankar} {et~al.}(2016){Shankar}, {Calderone}, {Knigge}, {Matthews},
  {Buckland}, {Hryniewicz}, {Sivakoff}, {Dai}, {Richardson}, {Riley}, {Gray},
  {La Franca}, {Altamirano}, {Croston}, {Gandhi}, {H{\"o}nig}, {McHardy}, \&
  {Middleton}}]{Shankar2016}
{Shankar}, F., {Calderone}, G., {Knigge}, C., {et~al.} 2016, \apjl, 818, L1

\bibitem[{{Shappee} {et~al.}(2014){Shappee}, {Prieto}, {Grupe}, {Kochanek},
  {Stanek}, {De Rosa}, {Mathur}, {Zu}, {Peterson}, {Pogge}, {Komossa}, {Im},
  {Jencson}, {Holoien}, {Basu}, {Beacom}, {Szczygie{\l}}, {Brimacombe},
  {Adams}, {Campillay}, {Choi}, {Contreras}, {Dietrich}, {Dubberley},
  {Elphick}, {Foale}, {Giustini}, {Gonzalez}, {Hawkins}, {Howell}, {Hsiao},
  {Koss}, {Leighly}, {Morrell}, {Mudd}, {Mullins}, {Nugent}, {Parrent},
  {Phillips}, {Pojmanski}, {Rosing}, {Ross}, {Sand}, {Terndrup}, {Valenti},
  {Walker}, \& {Yoon}}]{Shappee2014}
{Shappee}, B.~J., {Prieto}, J.~L., {Grupe}, D., {et~al.} 2014, \apj, 788, 48

\bibitem[{{Shields}(1978)}]{Sheilds1978}
{Shields}, G.~A. 1978, \nat, 272, 706

\bibitem[{{Skielboe} {et~al.}(2015){Skielboe}, {Pancoast}, {Treu}, {Park},
  {Barth}, \& {Bentz}}]{Skielboe2015}
{Skielboe}, A., {Pancoast}, A., {Treu}, T., {et~al.} 2015, \mnras, 454, 144

\bibitem[{{Starkey} {et~al.}(2017){Starkey}, {Horne}, {Fausnaugh}, {Peterson},
  {Bentz}, {Kochanek}, {Denney}, {Edelson}, {Goad}, {De Rosa}, {Anderson},
  {Ar{\'e}valo}, {Barth}, {Bazhaw}, {Borman}, {Boroson}, {Bottorff}, {Brandt},
  {Breeveld}, {Cackett}, {Carini}, {Croxall}, {Crenshaw}, {Dalla Bont{\`a}},
  {De Lorenzo-C{\'a}ceres}, {Dietrich}, {Efimova}, {Ely}, {Evans},
  {Filippenko}, {Flatland}, {Gehrels}, {Geier}, {Gelbord}, {Gonzalez},
  {Gorjian}, {Grier}, {Grupe}, {Hall}, {Hicks}, {Horenstein}, {Hutchison},
  {Im}, {Jensen}, {Joner}, {Jones}, {Kaastra}, {Kaspi}, {Kelly}, {Kennea},
  {Kim}, {Kim}, {Klimanov}, {Korista}, {Kriss}, {Lee}, {Leonard}, {Lira},
  {MacInnis}, {Manne-Nicholas}, {Mathur}, {McHardy}, {Montouri}, {Musso},
  {Nazarov}, {Norris}, {Nousek}, {Okhmat}, {Pancoast}, {Parks}, {Pei}, {Pogge},
  {Pott}, {Rafter}, {Rix}, {Saylor}, {Schimoia}, {Schn{\"u}lle}, {Sergeev},
  {Siegel}, {Spencer}, {Sung}, {Teems}, {Turner}, {Uttley}, {Vestergaard},
  {Villforth}, {Weiss}, {Woo}, {Yan}, {Young}, {Zheng}, \& {Zu}}]{Starkey2017}
{Starkey}, D., {Horne}, K., {Fausnaugh}, M.~M., {et~al.} 2017, \apj, 835, 65

\bibitem[{{Starkey} {et~al.}(2016){Starkey}, {Horne}, \&
  {Villforth}}]{Starkey2016}
{Starkey}, D.~A., {Horne}, K., \& {Villforth}, C. 2016, \mnras, 456, 1960

\bibitem[{{Suganuma} {et~al.}(2006){Suganuma}, {Yoshii}, {Kobayashi},
  {Minezaki}, {Enya}, {Tomita}, {Aoki}, {Koshida}, \&
  {Peterson}}]{Suganuma2006}
{Suganuma}, M., {Yoshii}, Y., {Kobayashi}, Y., {et~al.} 2006, \apj, 639, 46

\bibitem[{{Telfer} {et~al.}(2002){Telfer}, {Zheng}, {Kriss}, \&
  {Davidsen}}]{Telfer2002}
{Telfer}, R.~C., {Zheng}, W., {Kriss}, G.~A., \& {Davidsen}, A.~F. 2002, \apj,
  565, 773

\bibitem[{{Tody}(1986)}]{iraf}
{Tody}, D. 1986, in \procspie, Vol. 627, Instrumentation in astronomy VI, ed.
  D.~L. {Crawford}, 733

\bibitem[{{Troyer} {et~al.}(2016){Troyer}, {Starkey}, {Cackett}, {Bentz},
  {Goad}, {Horne}, \& {Seals}}]{Troyer2016}
{Troyer}, J., {Starkey}, D., {Cackett}, E.~M., {et~al.} 2016, \mnras, 456, 4040

\bibitem[{{Turner} {et~al.}(2006){Turner}, {Miller}, {George}, \&
  {Reeves}}]{Turner2006}
{Turner}, T.~J., {Miller}, L., {George}, I.~M., \& {Reeves}, J.~N. 2006, \aap,
  445, 59

\bibitem[{{Uttley} {et~al.}(2003){Uttley}, {Edelson}, {McHardy}, {Peterson}, \&
  {Markowitz}}]{Uttley2003}
{Uttley}, P., {Edelson}, R., {McHardy}, I.~M., {Peterson}, B.~M., \&
  {Markowitz}, A. 2003, \apjl, 584, L53

\bibitem[{{Valenti} {et~al.}(2015){Valenti}, {Sand}, {Barth}, {Horne}, {Treu},
  {Raganit}, {Boroson}, {Crawford}, {Pancoast}, {Pei}, {Romero-Colmenero},
  {Villforth}, \& {Winkler}}]{Valenti2015}
{Valenti}, S., {Sand}, D.~J., {Barth}, A.~J., {et~al.} 2015, \apjl, 813, L36

\bibitem[{van~der Walt {et~al.}(2011)van~der Walt, Colbert, \&
  Varoquaux}]{numpy}
van~der Walt, S., Colbert, S.~C., \& Varoquaux, G. 2011, Computing in Science
  {\&} Engineering, 13, 22

\bibitem[{{Vanden Berk} {et~al.}(2001){Vanden Berk}, {Richards}, {Bauer},
  {Strauss}, {Schneider}, {Heckman}, {York}, {Hall}, {Fan}, {Knapp},
  {Anderson}, {Annis}, {Bahcall}, {Bernardi}, {Briggs}, {Brinkmann}, {Brunner},
  {Burles}, {Carey}, {Castander}, {Connolly}, {Crocker}, {Csabai}, {Doi},
  {Finkbeiner}, {Friedman}, {Frieman}, {Fukugita}, {Gunn}, {Hennessy},
  {Ivezi{\'c}}, {Kent}, {Kunszt}, {Lamb}, {Leger}, {Long}, {Loveday}, {Lupton},
  {Meiksin}, {Merelli}, {Munn}, {Newberg}, {Newcomb}, {Nichol}, {Owen}, {Pier},
  {Pope}, {Rockosi}, {Schlegel}, {Siegmund}, {Smee}, {Snir}, {Stoughton},
  {Stubbs}, {SubbaRao}, {Szalay}, {Szokoly}, {Tremonti}, {Uomoto}, {Waddell},
  {Yanny}, \& {Zheng}}]{VandenBerk2001}
{Vanden Berk}, D.~E., {Richards}, G.~T., {Bauer}, A., {et~al.} 2001, \aj, 122,
  549

\bibitem[{{Vazquez} {et~al.}(2015){Vazquez}, {Galianni}, {Richmond},
  {Robinson}, {Axon}, {Horne}, {Almeyda}, {Fausnaugh}, {Peterson}, {Bottorff},
  {Gallimore}, {Eltizur}, {Netzer}, {Storchi-Bergmann}, {Marconi}, {Capetti},
  {Batcheldor}, {Buchanan}, {Stirpe}, {Kishimoto}, {Packham}, {Perez},
  {Tadhunter}, {Upton}, \& {Estrada-Carpenter}}]{Vazquez2015}
{Vazquez}, B., {Galianni}, P., {Richmond}, M., {et~al.} 2015, \apj, 801, 127

\bibitem[{{Veilleux} \& {Osterbrock}(1987)}]{Veilleux1987}
{Veilleux}, S., \& {Osterbrock}, D.~E. 1987, \apjs, 63, 295

\bibitem[{{Weedman}(1977)}]{Weedman1977}
{Weedman}, D.~W. 1977, \araa, 15, 69

\bibitem[{{Welsh}(1999)}]{Welsh1999}
{Welsh}, W.~F. 1999, \pasp, 111, 1347

\bibitem[{{White} \& {Peterson}(1994)}]{White1994}
{White}, R.~J., \& {Peterson}, B.~M. 1994, \pasp, 106, 879

\bibitem[{{Wilhite} {et~al.}(2005){Wilhite}, {Vanden Berk}, {Kron},
  {Schneider}, {Pereyra}, {Brunner}, {Richards}, \& {Brinkmann}}]{Whilhite2005}
{Wilhite}, B.~C., {Vanden Berk}, D.~E., {Kron}, R.~G., {et~al.} 2005, \apj,
  633, 638

\bibitem[{{Zu} {et~al.}(2013){Zu}, {Kochanek}, {Koz{\l}owski}, \&
  {Udalski}}]{Zu2013}
{Zu}, Y., {Kochanek}, C.~S., {Koz{\l}owski}, S., \& {Udalski}, A. 2013, \apj,
  765, 106

\bibitem[{{Zu} {et~al.}(2011){Zu}, {Kochanek}, \& {Peterson}}]{Zu2011}
{Zu}, Y., {Kochanek}, C.~S., \& {Peterson}, B.~M. 2011, \apj, 735, 80

\end{thebibliography}

\section*{Appendix}
\begin{figure*}
\includegraphics[width=\textwidth]{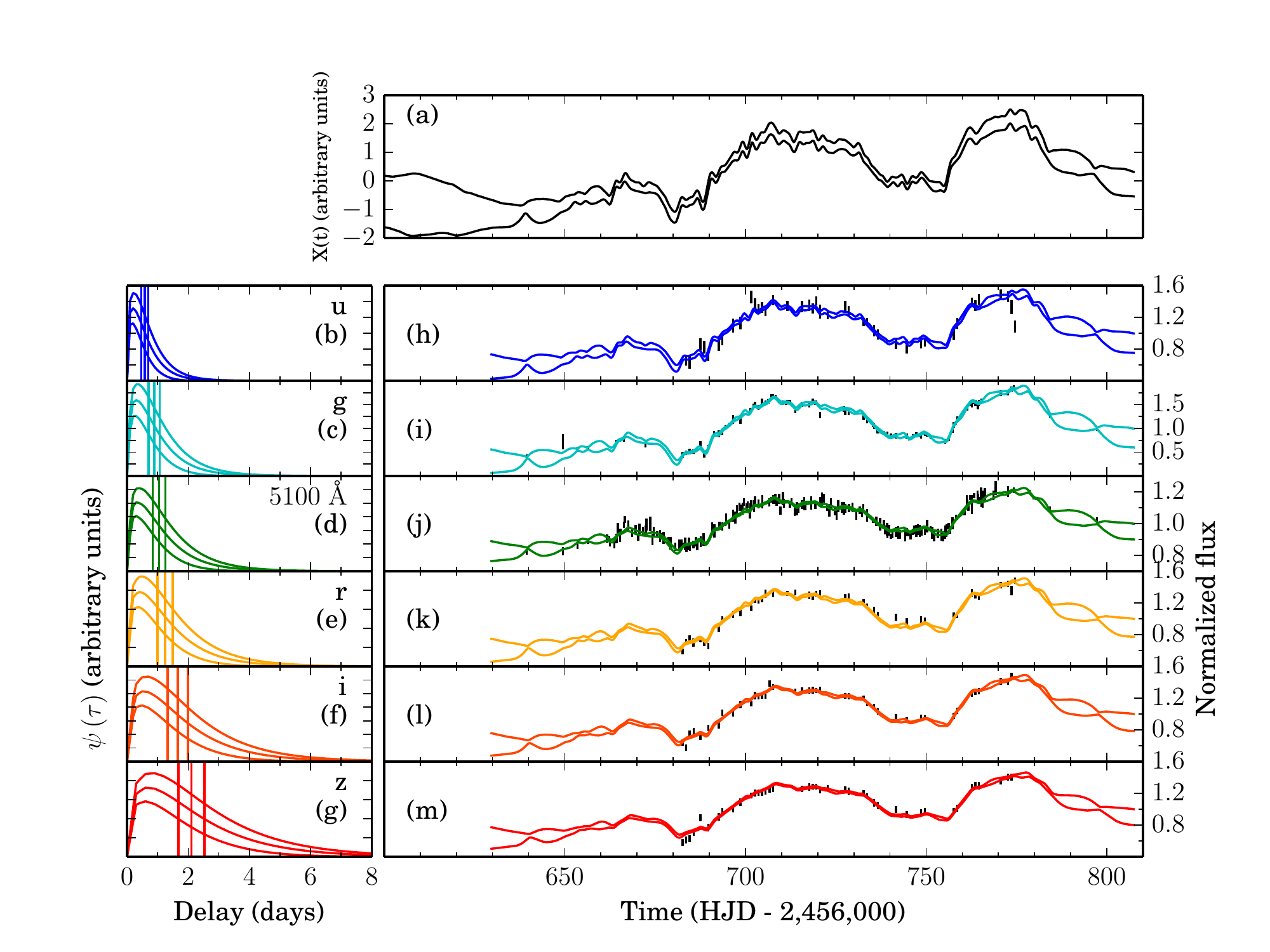}
\caption{CREAM fits to the {\it u} through {\z}-band light curves for
  MCG+08-11-011. Panel A shows the inferred driving light curve for
  the disk continuum variations. Panels b--g show the inferred
  transfer functions where the middle, lower and upper curves
  correspond to the mean and 1$\sigma$ uncertainties. The vertical
  lines denote the mean lags. Panels h--m show the light curves and
  uncertainty envelopes.\label{fig:mcg0811_cream1}}
\end{figure*}

\begin{figure*}
\includegraphics[width=\textwidth]{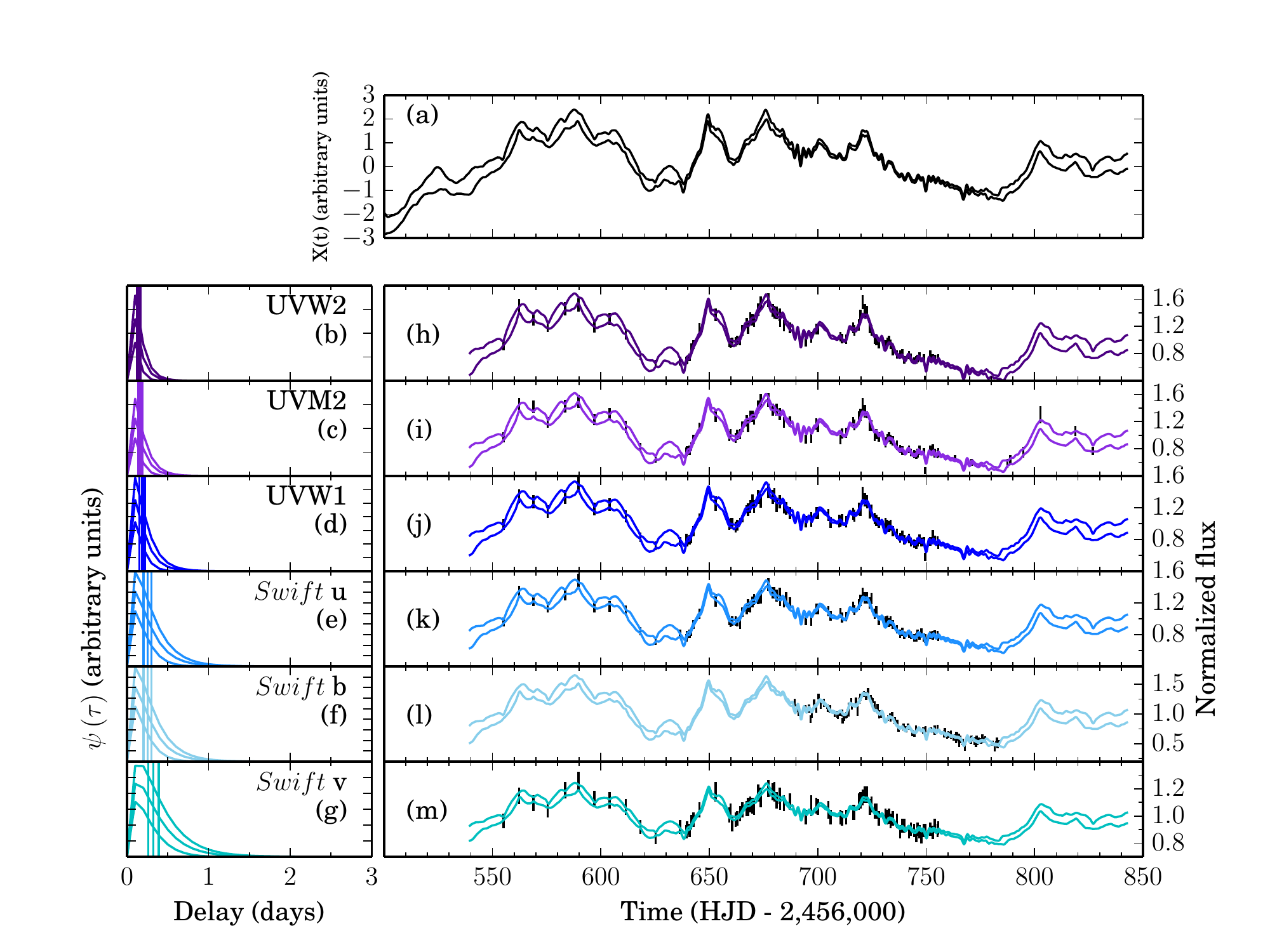}
\caption{Same as Figure\,\ref{fig:mcg0811_cream1} but for {\it Swift}
  data of NGC\,2617.\label{fig:n2617_cream1}}
\end{figure*}

\begin{figure*}
\includegraphics[width=\textwidth]{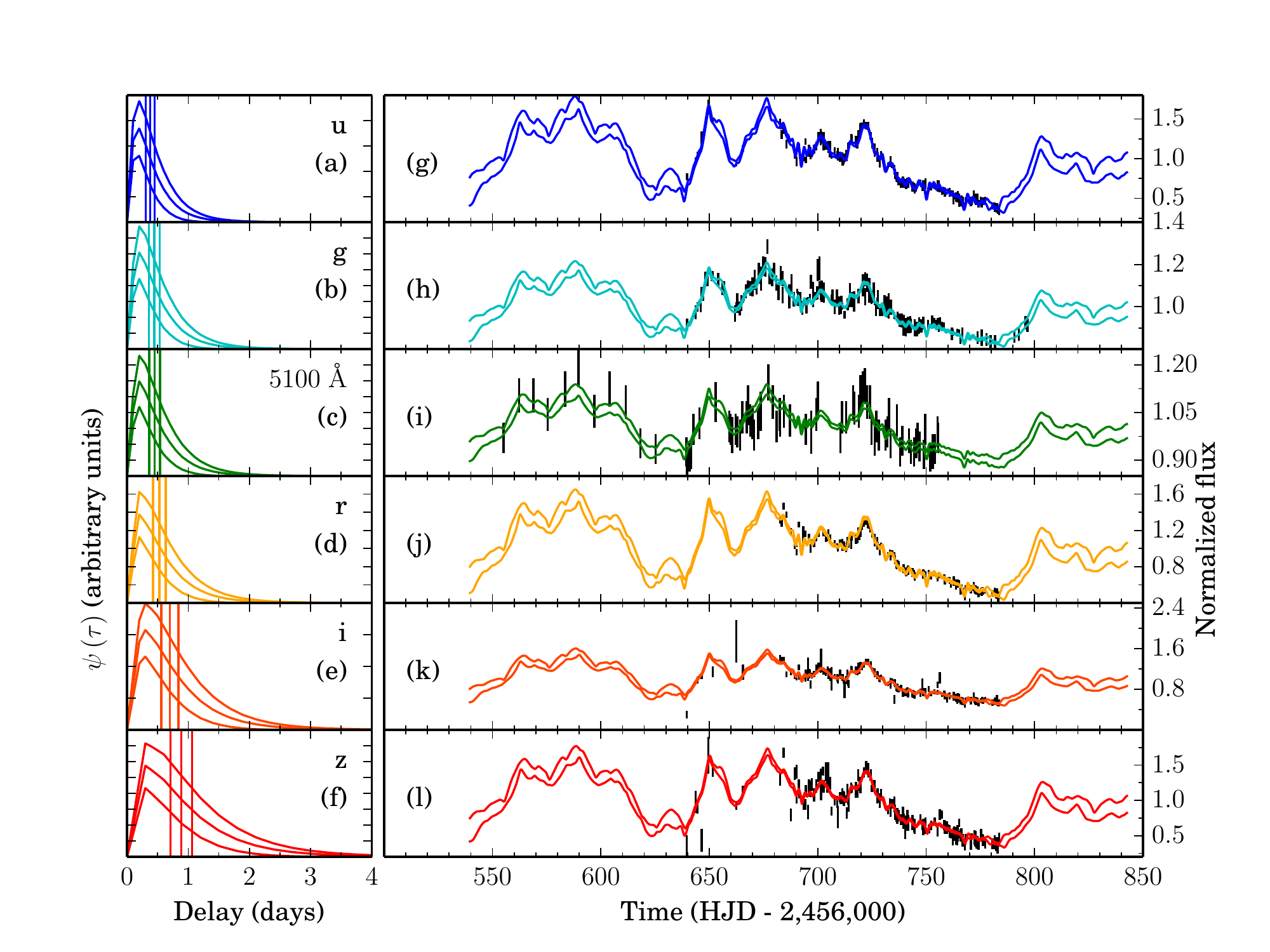}
\caption{Same as Figure\,\ref{fig:mcg0811_cream1} but for ground-based
  data of NGC\,2617.\label{fig:n2617_cream}}
\end{figure*}

\end{document}